\documentclass[a4paper,11pt]{article}
\usepackage{jcappub}
\usepackage{amssymb}
\usepackage{graphicx}
\usepackage{amsmath}
\usepackage{hyperref}
\usepackage{subfigure}
\usepackage{ulem}

\title{\boldmath The energy budget of cosmological first-order phase transitions beyond the bag equation of state}

\author[1]{Shao-Jiang Wang,}
\author[1,2]{Zi-Yan Yuwen}

\affiliation[1]{CAS Key Laboratory of Theoretical Physics, Institute of Theoretical Physics, Chinese Academy of Sciences, Beijing 100190, China}
\affiliation[2]{School of Physical Sciences, University of Chinese Academy of Sciences (UCAS), Beijing 100049, China}

\emailAdd{schwang@itp.ac.cn}
\emailAdd{yuwenziyan@itp.ac.cn (corresponding author)}

\abstract{The stochastic gravitational-wave backgrounds (SGWBs) from the cosmological first-order phase transitions (FOPTs) serve as a promising probe for the new physics beyond the standard model of particle physics. When most of the bubble walls collide with each other long after they had reached the terminal wall velocity, the dominated contribution to the SGWBs comes from the sound waves characterized by the efficiency factor of inserting the released vacuum energy into the bulk fluid motions. However, the previous works of estimating this efficiency factor have only considered the simplified case of the constant sound velocities in both symmetric and broken phases, either for the bag model with equal sound velocities or $\nu$-model with different sound velocities in the symmetric and broken phases, which is unrealistic from a viewpoint of particle physics. In this paper, we propose to solve the fluid EoM with an iteration method when taking into account the sound-velocity variation across the bubble wall for a general and realistic equation of state (EoS) beyond the simple bag model and $\nu$-model. We have found a suppression effect for the efficiency factor of bulk fluid motions, though such a suppression effect could be negligible for the strong FOPT, in which case the previous estimation from a bag EoS on the efficiency factor of bulk fluid motions still works as a good approximation.}

\begin{document}
\maketitle
\flushbottom

\section{Introduction}\label{sec:introduction}

Our current Universe is known in a symmetry-broken phase. Although the standard model (SM) of particle physics favors no strong evidence for the first-order phase transitions (FOPTs) in the early Universe but simply crossover transitions accroding to the lattice simulation results, the search for the FOPTs~\cite{Mazumdar:2018dfl,Hindmarsh:2020hop,Caldwell:2022qsj} is nevertheless welcome and well-motivated for some new physics beyond the SM~\cite{Cai:2017cbj,Bian:2021ini}, such as electroweak baryogenesis~\cite{Cohen:1990it,Cohen:1990py,Cohen:1993nk,Cohen:1994ss,Cohen:2012zza}, primordial magnetic field 
\cite{Hogan:1983zz,Quashnock:1988vs,Vachaspati:1991nm,Cheng:1994yr}, 
primordial black holes 
\cite{Hawking:1982ga,Kodama:1982sf,Moss:1994iq,Liu:2021svg,Hashino:2021qoq,Baker:2021nyl,Baker:2021sno,Kawana:2021tde,Huang:2022him,Marfatia:2021hcp}, 
stochastic gravitational waves backgrounds (SGWBs) 
\cite{Witten:1984rs,Hogan:1986qda,Kosowsky:1991ua,Kosowsky:1992rz,Kosowsky:1992vn,Kamionkowski:1993fg}, and so on. In particular, the peak frequency of the SGWB from a FOPT at the electroweak scale would fall into the frequency bands of LISA~\cite{Armano:2016bkm,Audley:2017drz}, Taiji~\cite{Hu:2017mde,Ruan:2018tsw,Taiji-1}, and TianQin~\cite{TianQin:2015yph,Luo:2020bls,TianQin:2020hid}, while a FOPT at the QCD scale would produce SGWB signals in the frequency regimes of the Pulsar Timing Array (PTA) and Square Kilometre Array (SKA). See~\cite{Binetruy:2012ze,Caprini:2015zlo,Weir:2017wfa,Caprini:2019egz} for recent reviews on SGWBs from the FOPTs.

The FOPT proceeds with stochastic nucleations of true vacuum bubbles in the background of the false vacuum if the false and true vacua are separated by a potential barrier in the effective potential. The nucleated bubbles then expand under the driving force given by the difference of the effective potential energy density between the false and true vacua. However, the bubble expansion is also subjected to the backreaction force~\cite{Wang:2022txy} from the thermal plasma fluid, which consists of the thermal force from the temperature variation around the bubble wall and the friction force from the non-equilibrium effect in the vicinity of the bubble wall. When the total backreaction force could eventually balances the driving force, the bubble expansion is of non-runaway type with some temrminal wall velocity, otherwise, it is a runaway expansion approaching to the speed of light, which is considered to be unrealistic according to the recent debates~\cite{Bodeker:2009qy,Bodeker:2017cim,Hoeche:2020rsg,Gouttenoire:2021kjv,Mancha:2020fzw,Vanvlasselaer:2020niz,Balaji:2020yrx,Ai:2021kak,Dorsch:2021nje,DeCurtis:2022hlx,Bea:2021zsu,Bigazzi:2021ucw}. For the runaway expansion or even the non-runaway expansion but with most of the bubble walls colliding with each other before approaching to the terminal wall velocity, the main contribution to the SGWBs comes from the bubble wall collisions~\cite{Kosowsky:1991ua,Kosowsky:1992rz,Kosowsky:1992vn,Kamionkowski:1993fg,Huber:2008hg}, in which case the efficiency factor $\kappa_\phi$ for inserting the released vacuum energy density into the kinetic energy of bubble walls is given in Ref.~\cite{Cai:2020djd}. Otherwise, for non-runaway expansion with most of the bubble walls colliding with each other long after approaching to the terminal wall velocity, the main contribution to the SGWBs comes from the sound waves~\cite{Hindmarsh:2013xza,Hindmarsh:2015qta,Hindmarsh:2017gnf,Cutting:2019zws}, in which case the efficiency factor $\kappa_v$ for inserting the released vacuum energy density into the bulk fluid motions is given in Ref.~\cite{Espinosa:2010hh} for a bag equation of state (EoS).

The bag EoS assumes no particle receiving a mass comparable to the background temperature during the FOPT. Since the heavy particles contribute to the effective potential in an exponentially suppressed manner, this bag EoS assumption simply takes into account the light particles alone in the broken phase as a collection of the thermal gas. If the symmetric phase is further assumed to be the thermal gas, the bag EoS then admits the sound velocity $c_s^2=1/3$ throughout the symmetric and broken phases if their corresponding vacuum energy densities $V_0(\phi_\pm)=\epsilon_\pm$ are also assumed to be constants. In short, the bag EoS assumes $p_\pm=c_s^2a_\pm T^4-\epsilon_\pm$ and $\rho_\pm=a_\pm T^4+\epsilon_\pm$ in the symmetric and broken phases. To go beyond the bag EoS, one can break any of above assumptions. For example, if some particles receive masses comparable to the transition temperature in the broken phase, the sound velocity in the broken phase $c_{s,-}^2<1/3$ could be different from that ($c_{s,+}^2\equiv1/3$) in the symmetric phase. Therefore, a simplified model called $\nu$-model was proposed in Ref.~\cite{Leitao:2014pda} for the most general EOS with a constant sound velocity by assuming $p_\pm=c_{s,\pm}^2a_\pm T^{\nu_\pm}-\epsilon_\pm$ and $\rho_\pm=a_\pm T^{\nu_\pm}+\epsilon_\pm$ with the sound velocities $c_{s,\pm}^2=\mathrm{d}p_\pm/\mathrm{d}\rho_\pm=1/(\nu_\pm-1)$, where the efficiency factor $\kappa_v$ of bulk fluid motions is estimated for a planar bubble wall expansion. The above $\nu$-model was further studied in Refs.~\cite{Giese:2020rtr,Giese:2020znk} with slightly different but equivalent form $p_\pm=\frac13a_\pm T^{\nu_\pm}-\epsilon_\pm$ and $\rho_\pm=\frac13a_\pm(\nu_\pm-1)T^{\nu_\pm}+\epsilon_\pm$ with the sound velocities $c_{s,\pm}^2=\mathrm{d}p_\pm/\mathrm{d}\rho_\pm=1/(\nu_\pm-1)$, where a pseudotrace $\alpha_{\bar{\theta}}$ is introduced to characterize the strength factor so that a model-independent approach can be achieved to express $\kappa_v$ solely in terms of $c_{s,\pm}$ and $\alpha_{\bar{\theta}}$ for the detonation~\cite{Giese:2020rtr}, deflagration and hybrid~\cite{Giese:2020znk} types of the bubble expansion. Similar discussion~\cite{Wang:2020nzm} appeared right after~\cite{Giese:2020znk} with a specific illustration from the Higgs sextic effective model. The effect of $\nu$-model on the GW spectrum was also studied in Ref.~\cite{Wang:2021dwl} for the sound shell model.

However, the hydrodynamic solutions from the fluid equation of motion (EoM) of the bubble expansion render spatially dependent fluid velocity, enthalpy, and temperature, which in turn would also give rise to a spatially inhomogeneous sound velocity profile in both symmetric and broken phases not only far from but also near the bubble wall. Since the fluid EoM itself already contains the sound velocity $c_s^2\equiv\mathrm{d}p(T)/\mathrm{d}\rho(T)$ as a function of the temperature profile, the $\nu$-model is thus not general enough for the realistic particle physics models. Therefore, in this paper, we assume the most general EoS to date and then iteratively solve the fluid EoM consistently with a spatially dependent sound velocity profile. The efficiency factor of bulk fluid motions is obtained in a way that can be directly compared to the case with a bag EoS, where a suppression effect is illustrated that can be neglected for a strong FOPT.

The outline of this paper is as follows: In Section~\ref{sec:wibag}, we briefly review the thermodynamics in~\ref{subsec:thermodynamics_wibag} and hydrodynamics in~\ref{subsec:hydrodynamics_wibag} for the case with a bag EoS, and then solve for the velocity, enthalpy and temperature profiles for three expansion modes in~\ref{subsec:expansion_wibag}. Next, we go beyond bag EoS in Section~\ref{sec:wobag}, where our general EoS is given in~\ref{subsec:thermodynamics_wobag}. After iteratively solving the fluid EoM in~\ref{subsec:hydrodynamics_wobag} for three expansion modes in~\ref{subsec:expansion_wobag}, we finally obtain the efficiency factor in~\ref{subsec:efficiency}. The Section~\ref{sec:con} is devoted to conclusions and discussions.

\section{Bubble expansion with a bag equation of state}\label{sec:wibag}

In this section, we briefly review the thermodynamics and hydrodynamics of the bubble expansion with a bag EoS for three types of expansion modes as detailed in Ref.~\cite{Espinosa:2010hh}.

\subsection{Thermodynamics}\label{subsec:thermodynamics_wibag}

The starting point is the bag EoS ansatz for the pressure and energy density of forms
\begin{align}
    p_\pm=\frac{1}{3}a_\pm T^4 - \epsilon_\pm, \quad
    \rho_\pm=a_\pm T^4 + \epsilon_\pm
\end{align}
in the symmetric and broken phases labeled by the plus and minus subscripts, respectively. Here $\epsilon_\pm\equiv V_0(\phi_\pm)$ is evaluated from the temperature-independent part of the total effective potential. We have assumed a negligible temperature dependence in the vacuum expectation value $\phi_-$, which is usually the case for the most of the particle physics models with a FOPT. $a_\pm$ is evaluated from the number $g_i$ of the relativistic degrees of freedom of species $i$ by
\begin{align}
    a = \frac{\pi^2}{30} \left( \sum_{\mathrm{light\,Boson}}g_i + \frac{7}{8}\sum_{\mathrm{light\,Fermion}}g_j \right). \label{eq:a}
\end{align}
Usually one has $a_+>a_-$ since the symmetric phase has more light degrees of freedom than broken phase. This bag EoS ansatz is motivated from identifying the free energy density of the scalar-plasma system as the total effective potential with the leading-order thermal correction,
\begin{align}
    \mathcal{F}(\phi,T) = V_{\mathrm{eff}}(\phi,T) \approx V_{0}(\phi) - \frac{1}{3}a T^4. \label{eq:Veff0order}
\end{align}
Therefore, the pressure, energy density and enthalpy follow from the simple thermodynamical relations as
\begin{align}
    p =& -\mathcal{F} = -V_{\mathrm{eff}} 
    = -V_{0}(\phi) + \frac{1}{3}a T^4 , \label{eq:pibag} \\
    \rho =& T\partial_T p - p = V_{\mathrm{eff}} - T\partial_T V_{\mathrm{eff}} 
    = V_{0}(\phi) + a T^4 , \label{eq:rhoibag}\\
    w =& \rho + p = -T\partial_T V_{\mathrm{eff}} = \frac{4}{3} a T^4,
\end{align}
respectively. Note here that the summation in Eq.~\eqref{eq:a} only covers light particles with their masses much smaller than the background temperature, while the contributions from heavy particles to the effective potential are exponentially suppressed. As a result, only particles with their mass comparable to the background temperature would lead to a deviation from the bag EoS, which is omited in this section but included in the next section when we go beyond the bag EoS.

\subsection{Hydrodynamics}\label{subsec:hydrodynamics_wibag}

The hydrodynamics of the system can be described by the conservation equation of the total energy-momentum tensor in the bulk fluid and at the bubble wall interface, where the total energy-momentum tensor for the scalar-plasma system is usually approximated as a perfect fluid,
\begin{align}
    T^{\mu\nu}=(\rho+p)u^\mu u^\nu+p\eta^{\mu\nu} = w u^\mu u^\nu+p\eta^{\mu\nu},
\end{align}
given the pressure $p$, enthalpy $w$, and the fluid four-velocity $u^\mu=\gamma(v)(1,\vec{v})$ in the background plasma frame with the Lorentz factor $\gamma(v)=1/\sqrt{1-v^2}$. In what follows, we will first derive the fluid EoMs from the conservation of the energy-momentum tensor in the bulk fluid, and then obtain the junction conditions from the conservation of the energy-momentum tensor at the bubble wall interface as boundary conditions for the fluid EoM.

The fluid EoMs are derived from projecting the conservation equation of the energy-momentum tensor $\nabla_\mu T^{\mu\nu} = 0$ parallel and perpendicular to the bulk flow direction by $u_\nu\nabla_\mu T^{\mu\nu}=0$ and $\tilde{u}_\nu\nabla_\mu T^{\mu\nu}=0$ with $u^\mu=\gamma(v)(1,\vec{v})$ and $\tilde{u}^\mu=\gamma(v)(v,\vec{v}/v)$ satisfying 
\begin{align}
    u_\mu u^\mu=-1, \quad
    \tilde{u}_\mu\tilde{u}^\mu=1, \quad
    u_\nu\nabla_\mu u^\nu=0, \quad
    \tilde{u}_\mu u^\mu=0,
\end{align}
and the resulted equations
\begin{align}
    w\nabla_\mu u^\mu + u^\mu\nabla_\mu\rho =& 0,\\
    w\tilde{u}_\nu u^\mu\nabla_\mu u^\nu + \tilde{u}^\mu\nabla_\mu p =& 0.
\end{align}
At a finite temperature, the bubble is nucleated with the scalar profile given by the $O(3)$ bounce solution with a spherical symmetry, and the subsequent expansion is also assumed to be spherical so that the spherical coordinates are conveniently used, thus the fluid velocity and other thermodynamic quantities depend only on the temporal and radial coordinates $t$ and $r$, respectively, where the radial coordinate $r$ is the distance to the bubble center and the temporal coordinate $t$ is the time since the bubble nucleation. Furthermore, when the bubble reaches a steady expansion state long after the bubble nucleation, the initial size of the bubble can be assumed to be negligible, thus there is no characteristic distance scale for a steadily expanding bubble, and hence the fluid velocity and other thermodynamic quantities depend only on the dubbed self-similarity coordinate $\xi\equiv r/t$. Therefore, the above fluid EoMs can be re-written with the self-similarity coordinate as
\begin{align}
    (\xi-v)\frac{\partial_\xi\rho}{w}&=2\frac{v}{\xi}+\gamma^2(1-\xi v)\partial_\xi v,\\
    (1-\xi v)\frac{\partial_\xi p}{w}&=\gamma^2(\xi-v)\partial_\xi v,
    \label{eq:dp}
\end{align}
where $v(\xi)$ is the fluid velocity at $r=\xi t$ seen by an observer in the bubble center frame. After rearrange the above equations by division and summation via $c_s^2=\partial_\xi p/\partial_\xi\rho$ and $w=\rho+p$, the fluid EoMs are obtained as
\begin{align}
    2\frac{v}{\xi}&=\gamma^2(1-\xi v) \left(\frac{\mu(\xi,v)^2}{c_s^2}-1\right)\frac{\mathrm{d}v}{\mathrm{d}\xi},  \label{eq:EoMibag}\\
    \frac{\mathrm{d}w}{\mathrm{d}\xi}&=w\gamma^2\mu(\xi,v)\left(\frac{1}{c_s^2}+1\right)\frac{\mathrm{d}v}{\mathrm{d}\xi}, \label{eq:enthalpyibag}
\end{align}
where the abbreviation
\begin{align}
    \mu(\xi,v)=\frac{\xi-v}{1-\xi\cdot v}
\end{align}
is simply the fluid velocity seen from a frame moving with the velocity $\xi$. In particular, for a steady expansion with a terminal wall velocity $\xi_w$, the expressions $\mu(\xi_w, v)\equiv\bar{v}$ and $\mu(\xi_w,\bar{v})\equiv v$ are simply the Lorentz-boost transformation of the fluid velocity between the bubble wall frame (with a overbar symbol) and background plasma frame (without the overbar symbol).

The junction conditions are determined from the conservation equation of the energy-momentum tensor $\nabla_\mu T^{\mu\nu} = 0$  at the bubble wall interface, which is simply the continuity equations of the energy-momentum flow,
\begin{align}
    (T_+^{z \nu}-T_-^{z \nu})n_\nu=0, \quad (T_+^{t \nu}-T_-^{t \nu})n_\nu=0,
\end{align}
perpendicular to the bubble wall along the unit spatial vector $n_\mu = (0,0,0,1)$ of $z$ direction. After written in the bubble wall frame, the continuity equations become
\begin{align}
    w_+\bar{v}_+\bar{\gamma}_+^2&=w_-\bar{v}_-\bar{\gamma}_-^2,
    \label{eq:junc_wibag}\\
    w_+\bar{v}_+^2\bar{\gamma}_+^2+p_+&=w_-\bar{v}_-^2\bar{\gamma}_-^2+p_-,
\end{align}
which are the junction conditions relating the pressure, enthalpy, and fluid velocity across the bubble wall. After rearranging the junction conditions,
\begin{align}
    \bar{v}_+\bar{v}_- &= \frac{p_+-p_-}{\rho_+-\rho_-} = \frac{1-(1-3\alpha_+)r}{3-3(1+\alpha_+)r}, \label{eq:vbptimesvbm}\\
    \frac{\bar{v}_+}{\bar{v}_-} &= \frac{\rho_-+p_+}{\rho_++p_-} = \frac{3+(1-3\alpha_+)r}{1+3(1+\alpha_+)r},\label{eq:vbpdividevbm}
\end{align}
the fluid velocity can be solved as a function of thermodynamic quantities by
\begin{align}
    \bar{v}_+(\alpha_+,r)&=\sqrt{\frac{1-(1-3\alpha_+)r}{3-3(1+\alpha_+)r}\cdot\frac{3+(1-3\alpha_+)r}{1+3(1+\alpha_+)r}}, \label{eq:vbpibag}\\
    \bar{v}_-(\alpha_+,r)&=\sqrt{\left.\frac{1-(1-3\alpha_+)r}{3-3(1+\alpha_+)r}\right/\frac{3+(1-3\alpha_+)r}{1+3(1+\alpha_+)r}}, \label{eq:vbmibag}
\end{align}
for a bag EoS with the abbreviations 
\begin{align}
    \alpha_+=\frac{\Delta\epsilon}{a_+T_+^4}=\frac{4\Delta\epsilon}{3w_+}, \quad r=\frac{w_+}{w_-}=\frac{a_+ T_+^4}{a_-T_-^4}.
    \label{eq:def_r_wibag}
\end{align}

\subsection{Expansion modes}\label{subsec:expansion_wibag}

Solving the fluid EoM~\eqref{eq:EoMibag} with the junction conditions~\eqref{eq:junc_wibag}, the fluid velocity profile $v(\xi)$ can be obtained for three types of the expansion modes as shown shortly below in Fig.~\ref{fig:profilesibag}. With the solved fluid velocity profile, the enthalpy profile can also be obtained from Eq.~\eqref{eq:enthalpyibag} as
\begin{align}
    w(\xi)=w(\xi_0)\exp\left[\int_{v(\xi_0)}^{v(\xi)}\left(\frac{1}{c_s^2}+1\right)\gamma^2\mu(\xi,v)\mathrm{d}v\right].
    \label{eq:enthalpy_wibag}
\end{align}
The temperature profile can also be obtained from $\partial_\xi\ln T=\gamma^2\mu\partial_\xi v$ via~\eqref{eq:dp} as 
\begin{align}
    T(\xi)=T(\xi_0)\exp\left[\int_{v(\xi_0)}^{v(\xi)}\gamma^2\mu(\xi,v)\mathrm{d}v\right]. \label{eq:temperature}
\end{align}
Note that both the enthalpy and temperature profiles at the discontinuity interface $\xi_0$ from either the bubble wall or shockwave front should be specified from the junction conditions~\eqref{eq:junc_wibag} and~\eqref{eq:def_r_wibag}.

\begin{figure}
    \centering
    \subfigure[$v(\xi)$ for detonation]{
    \includegraphics[width=0.47\textwidth]{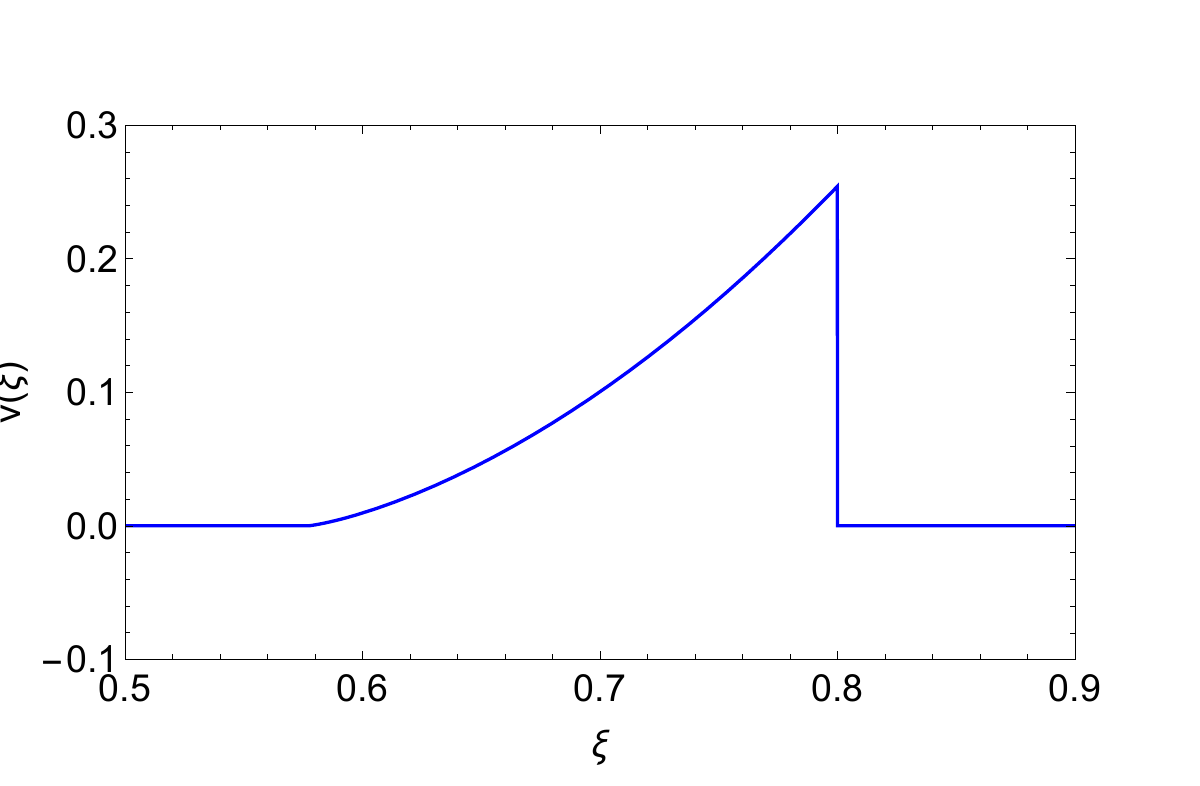}
    \label{fig:detonav_wibag}
    }
    \subfigure[$w(\xi)$ and $T(\xi)$ for detonation]{
    \includegraphics[width=0.47\textwidth]{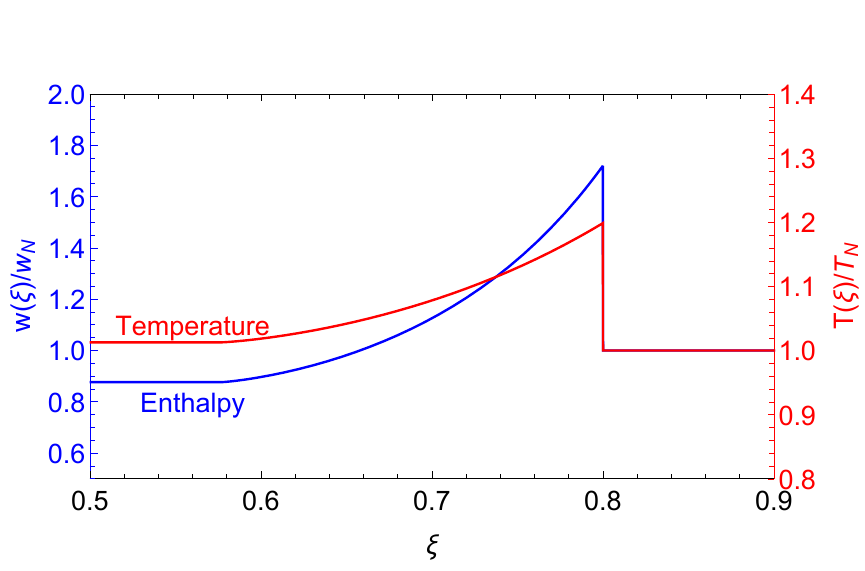}
    \label{fig:detonawt_wibag}
    }
    \subfigure[$v(\xi)$ for hybrid]{
    \includegraphics[width=0.47\textwidth]{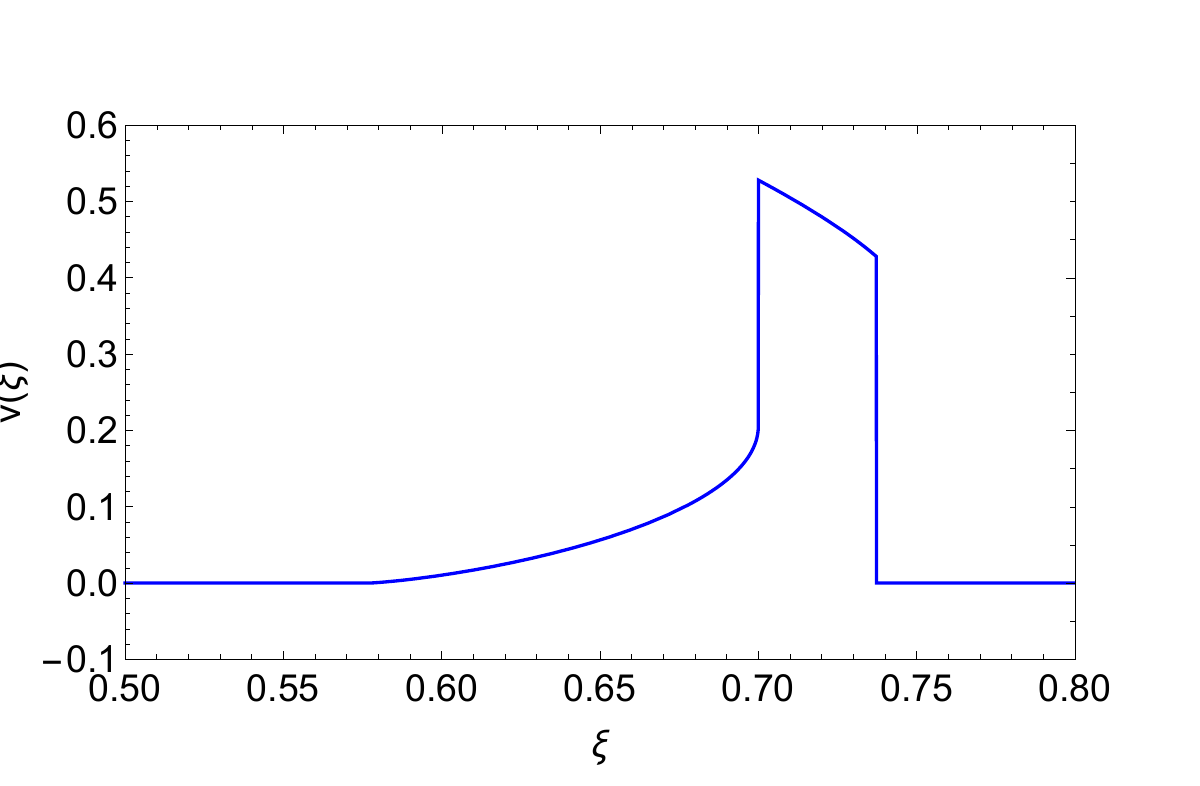}
    \label{fig:hybridv_wibag}
    }
    \subfigure[$w(\xi)$ and $T(\xi)$ for hybrid]{
    \includegraphics[width=0.47\textwidth]{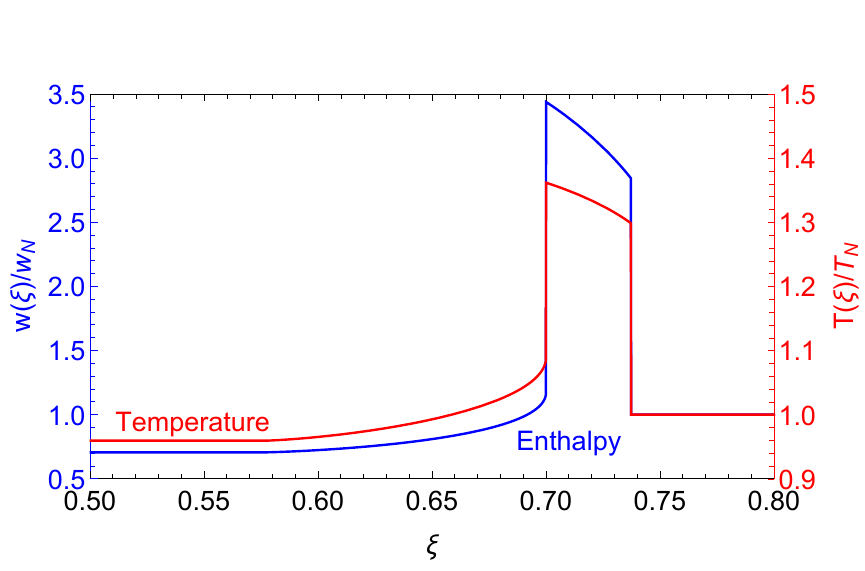}
    \label{fig:hybridwt_wibag}
    }
    \subfigure[$v(\xi)$ for deflagration]{
    \includegraphics[width=0.47\textwidth]{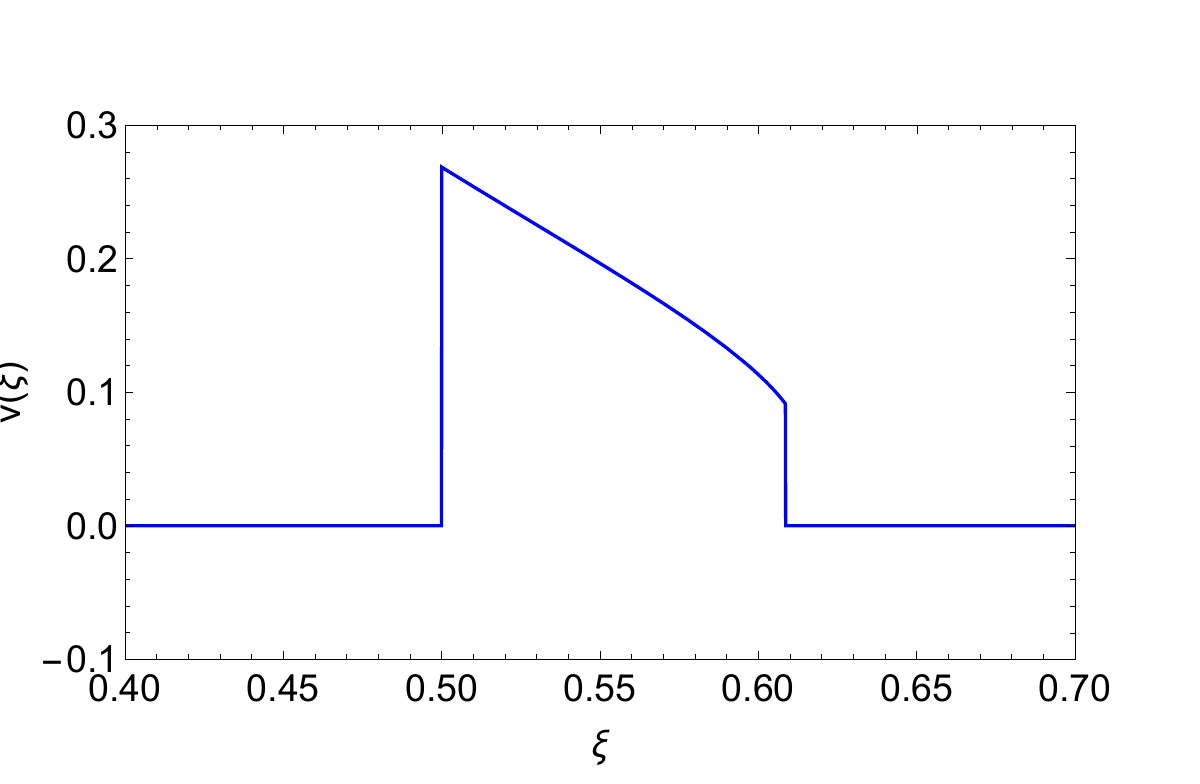}
    \label{fig:deflagv_wibag}
    }
    \subfigure[$w(\xi)$ and $T(\xi)$ for deflagration]{
    \includegraphics[width=0.47\textwidth]{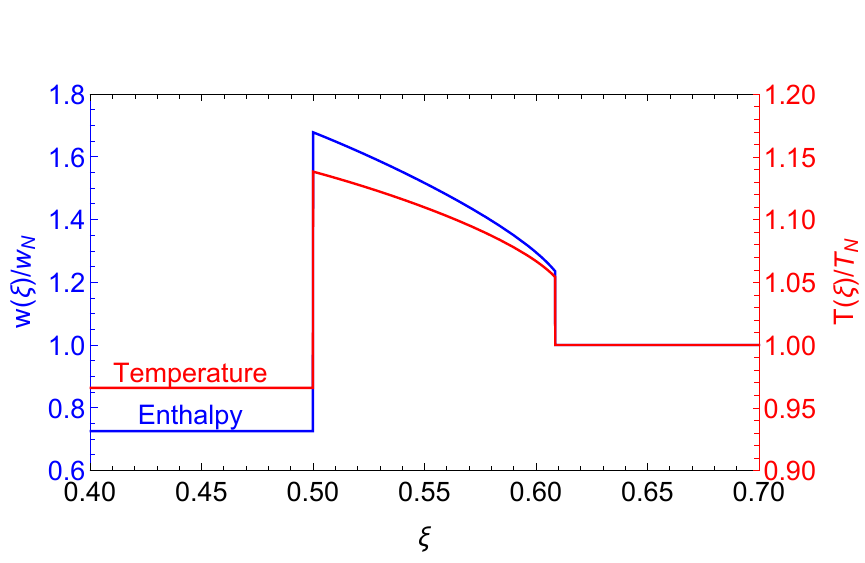}
    \label{fig:deflagwt_wibag}
    }
    \caption{The profiles of the fluid velocity (left column), enthalpy (right column in blue), and temperature (right column in red) for the bubble expansion modes of detonation (top row), hybrid (middle row), and deflagration (bottom row) types with a bag EoS are illustrated for a typical set of parameter choice with $\alpha_+=0.1$, $a_+/a_-=1.2$.}
    \label{fig:profilesibag}
\end{figure}

\subsubsection{Detonation}\label{subsec:detona_wibag}

The physically allowed detonation mode ($\bar{v}_+>\bar{v}_-$) is of the weak type with $\bar{v}_->c_s$, which can be achieved by a supersonic wall so that the wall-fluid interaction cannot propagate to the plasma fluid in the front of the bubble wall, that is, the fluid velocity in the front of the bubble wall remains static in the background plasma frame, $v_+ = 0 = \mu(\xi_w, \bar{v}_+)$, namely, $\bar{v}_+ = \xi_w$. Further using the expression of $\bar{v}_+$ from Eq.~\eqref{eq:vbpibag}, one can inversely solve for $r(\xi_w,\alpha_+)$, which, after inserted into Eq.~\eqref{eq:vbmibag}, gives rise to $\bar{v}_-(\alpha_+,r(\xi_w,\alpha_+))\equiv\bar{v}_-(\xi_w,\alpha_+)$ and hence $v_-=\mu(\xi_w,\bar{v}_-)$. Note that $\xi_w\equiv\bar{v}_+>\bar{v}_-(\xi_w,\alpha_+)$ immediately requires the supersonic wall velocity to be larger than the so-called Jouguet velocity $v_J$,
\begin{align}\label{eq:vJibag}
\xi_w>\frac{\sqrt{\alpha_+(2+3\alpha_+)}+1}{\sqrt{3}(1+\alpha_+)}\equiv v_J.
\end{align}
Having specified the boundary condition $(\xi_w^-,v_-)$ right after the bubble wall, the fluid velocity profile can be solved from the fluid EoM~\eqref{eq:EoMibag} as shown in Fig.~\ref{fig:detonav_wibag}.
To further specify the boundary conditions for the profiles of the enthalpy~\eqref{eq:enthalpy_wibag} and temperature~\eqref{eq:temperature}, one applies the junction condition 
$ w_+\bar{v}_+\bar{\gamma}_+^2=w_-\bar{v}_-\bar{\gamma}_-^2$ across the bubble wall with 
\begin{align}
    w_- = w(\xi_w^-), \quad w_+ = w_N, \quad \bar{v}_-=\bar{v}_-(\alpha_+,r), \quad \bar{v}_+ = \xi_w,
\end{align}
which gives rise to the enthalpy $w_-$ just behind the bubble wall as
\begin{align}
    w(\xi_w^-) = w_+\frac{\bar{v}_+ \gamma^2(\bar{v}_+)}{\bar{v}_- \gamma^2(\bar{v}_-)} = w_N \frac{\xi_w}{1-\xi_w^2} \frac{1-\bar{v}_-^2}{\bar{v}_-},
\end{align}
while the temperature $T_-$ just behind the bubble wall can be obtained from the definition of $r$ in~\eqref{eq:def_r_wibag} as
\begin{align}
    T_- = T_+ \left( r\frac{a_-}{a_+} \right)^{-1/4} =T_N \left( \frac{w_-}{w_+}\frac{a_+}{a_-} \right)^{1/4}.
\end{align}
Here the subscript $N$ will always denote for the asymptotic value far in the front of the bubble wall. The resulted enthalpy and temperature profiles for detonation are shown in Fig.~\ref{fig:detonawt_wibag}.

\subsubsection{Deflagration}\label{subsec:deflag_wibag}

The physically allowed deflagration mode ($\bar{v}_+<\bar{v}_-$) is also of the weak type with $\bar{v}_-<c_s$, which can be achieved by a subsonic wall with the compressive shockwave in the front of the bubble wall, while the wall-fluid interaction does not propagate to the plasma fluid at the back of the bubble wall, that is, the fluid velocity at the back of the bubble wall remains static in the background plasma frame, $v_- = 0 = \mu(\xi_w, \bar{v}_-)$, namely, $\bar{v}_- = \xi_w$. Similar to the detonation mode, further using the expression of $\bar{v}_-$ from Eq.~\eqref{eq:vbmibag}, one can also inversely solve for $r(\xi_w,\alpha_+)$, which, after inserted into Eq.~\eqref{eq:vbpibag}, gives rise to $\bar{v}_+(\alpha_+,r(\xi_w,\alpha_+))\equiv\bar{v}_+(\xi_w,\alpha_+)$ and hence $v_+=\mu(\xi_w,\bar{v}_+)$. Starting from the boundary condition $(\xi_w^+, v_+)$, the fluid velocity profile can be solved from the fluid EoM~\eqref{eq:EoMibag} forward in $\xi$ until arriving at the shockwave front where the fluid velocity jumps to zero. To further specify the boundary condition at the shockwave front, one simply applies the junction condition~\eqref{eq:vbptimesvbm} to a frame comoving with the shockwave front, $\bar{v}'_+ \bar{v}'_- = 1/3$ with $\alpha'_+=0$ at the shockwave front, where the prime symbol will always denote the quantity in the shockfront rest frame, and the associated plus and minus symbols correspond to the position right in the front and behind the shockwave front $\xi_{sh}$, respectively. Since the fluid velocity in the front of the shockwave front also remains static in the background plasma frame, $\bar{v}'_+ = \mu(\xi_{sh},0) = \xi_{sh}$, the fluid velocity $v_{sh}\equiv v(\xi_{sh}^-)$ right behind the shockwave front is therefore solved from
\begin{align}
    \xi_{sh}\cdot\mu(\xi_{sh},v_{sh}) = 1/3. \label{eq:shockfront}
\end{align}
With these two boundary conditions at the bubble wall $(\xi_w^+,v_+)$ and shockwave front $(\xi_{sh}^-,v_{sh})$, the fluid velocity profile solved from the fluid EoM~\eqref{eq:EoMibag} can be fixed as shown in Fig.~\ref{fig:deflagv_wibag}.

To further specify the boundary conditions for computing the enthalpy profile from Eq.~\eqref{eq:enthalpy_wibag}, one simply
applies the junction condition $w_-\bar{v}'_-\gamma^2(\bar{v}'_-)=w_+\bar{v}'_+\gamma^2(\bar{v}'_+)$ across the shockwave front with
\begin{align}
    w_- = w(\xi_{sh}^-), \quad w_+ = w_N, \quad \bar{v}'_-=\bar{v}'_-(\alpha'_+\equiv0,r'), \quad \bar{v}'_+ = \xi_{sh},
\end{align}
so that the enthalpy right behind the shockwave front reads
\begin{align}
    w(\xi_{sh}^-) = w_+\frac{\bar{v}'_+ \gamma^2(\bar{v}'_+)}{\bar{v}'_- \gamma^2(\bar{v}'_-)} = 
    w_N \frac{\xi_{sh}}{1-\xi_{sh}^2} \frac{1-\mu(\xi_{sh},v(\xi_{sh}))^2}{\mu(\xi_{sh},v(\xi_{sh}))}.
\end{align}
Having specified the boundary condition $(\xi_{sh}^-,w(\xi_{sh}^-))$ at the shockwave front, one can evolve the $w( \xi )$ backward in $\xi$ according to Eq.~\eqref{eq:enthalpy_wibag} until reaching the bubble wall at $(\xi_w^+, w(\xi_w^+)))$, where the boundary condition $ w_+\bar{v}_+\bar{\gamma}_+^2=w_-\bar{v}_-\bar{\gamma}_-^2$ at the bubble wall with
\begin{align}
    w_-=w(\xi_w^-), \quad w_+=w(\xi_w^+), \quad \bar{v}_-=\xi_w, \quad 
    \bar{v}_+=\mu(\xi_w,v(\xi_w)),
\end{align}
gives rise to the enthalpy right behind the bubble wall,
\begin{align}
    w(\xi_w^-) = w_+\frac{\bar{v}_+ \gamma^2(\bar{v}_+)}{\bar{v}_- \gamma^2(\bar{v}_-)} = w(\xi_w^+) \frac{1-\xi_w^2}{\xi_w} \frac{\mu(\xi_{w},v(\xi_{w}))}{1-\mu(\xi_{w},v(\xi_{w}))^2},
\end{align}

As for the boundary conditions when computing the temperature profile from Eq.~\eqref{eq:temperature}, one first notices that the temperature jump can be determined from the enthalpy jump by
\begin{align}
    \frac{a_+T_+^4}{a_-T_-^4} = \frac{w_+}{w_-}.
\end{align}
Since $a_+ = a_-$ across the shockwave front, the temperature jump at the shockwave front simply reads
\begin{align}
    T_- =  T_N \left( \frac{w(\xi_{sh}^-)}{w_N} \right)^{1/4}.
\end{align}
Evolving the temperature profile $T(\xi)$ backward in $\xi$ accroding to Eq.~\eqref{eq:temperature} until reaching the bubble wall at $(\xi_w^+,T(\xi_w^+))$, where the temperature jump right behind the bubble wall reads
\begin{align}
    T(\xi) \equiv T(\xi_w^-) = T(\xi_w^+) \left( \frac{w(\xi_w^-)}{w(\xi_w^+)} \frac{a_+}{a_-} \right)^{1/4},
\end{align}
The resulted enthalpy and temperature profiles for deflagration are shown in Fig.~\ref{fig:deflagwt_wibag}.

\subsubsection{Hybrid}\label{subsec:hybrid_wibag}

The hybrid expansion ($\bar{v}_+<\bar{v}_-=c_s$) occurs when the bubble wall moves faster than the sound velocity but smaller than the Jouguet velocity~\eqref{eq:vJibag} determined directly from the Jouguet detonation condition $\bar{v}_-=c_{s,-}$ (the sound velocity in the broken phase assuming to be $1/\sqrt{3}$ with a bag EoS). Since the hybrid mode contains both the shockwave in the front the bubble wall and rarefaction wave behind the bubble wall, the velocity profile is therefore divided into four pieces, each of which can be determined in a similar way as we did in Section~\ref{subsec:detona_wibag} and Section~\ref{subsec:deflag_wibag}. For a fixed $\alpha_+$, the fluid in the region $\xi>\xi_{sh}$ stays at rest in the background plasma frame, where boundary condition at $(\xi_{sh}^-,v(\xi_{sh}^-))$ is determined from $\xi_{sh}\mu(\xi_{sh},v(\xi_{sh}))=c_s^2$. For the fluid in the region $\xi_w<\xi<\xi_{sh}$ and $1/\sqrt{3}\equiv c_{s,-}<\xi<\xi_{w}$, the velocity profiles can be evolved via the fluid EoM~\eqref{eq:EoMibag} both forward and backward in $\xi$ from the boundary condition at $(\xi_w^\pm, v_\pm)$, where $v_\pm\equiv\mu(\xi_w,\bar{v}_\pm)$ can be obtained from $\bar{v}_\pm\equiv\bar{v}_\pm(\alpha_+,r(\alpha_+))$ with $r(\alpha_+)$ determined by the Jouguet condition $\bar{v}_-(\alpha_+,r)=c_s$. At last, the fluid in the region $\xi<c_{s,-}$ stays at rest. The resulted fluid velocity profile for the hybrid mode is shown in Fig.~\ref{fig:hybridv_wibag}.

Similar procedures also apply to the enthalpy and temperature profiles when specifying their boundary conditions at the bubble wall and shockwave front. In the front of the shockwave front, both the enthalpy and temperature stay at their asymptotic values. After applying the junction condition across the shockwave front, the enthalpy and temperature right behind the shorckwave front read
\begin{align}
    w(\xi_{sh}^-) = w_N \frac{\xi_{sh}}{1-\xi_{sh}^2} \frac{1-\mu(\xi_{sh},v(\xi_{sh}))^2}{\mu(\xi_{sh},v(\xi_{sh}))}, \quad
    T(\xi_{sh}^-) = T_N \left( \frac{w(\xi_{sh}^-)}{w_N} \right)^{1/4}.
\end{align}
Evolving the enthalpy and temperature profiles $w(\xi)$ and $T(\xi)$ backward in $\xi$ via Eq.~\eqref{eq:enthalpy_wibag} and~\eqref{eq:temperature} from the above boundary conditions at $(\xi_{sh}^-, w(\xi_{sh}^-))$ and $(\xi_{sh}^-, T(\xi_{sh}^-))$, respectively, one arrives at the enthalpy and temperature $w(\xi_w^+)$ and $T(\xi_w^+)$ right in the front of the bubble wall, which can be used to determine the enthalpy and temperature right behind the bubble wall as
\begin{align}
    w(\xi_w^-) = w(\xi_w^+) \frac{\bar{v}_+}{1-\bar{v}_+^2} \frac{1-(c_s^-)^2}{c_s^-},
    \quad
    T(\xi_w^-) = T(\xi_w^+) \left( \frac{w(\xi_w^-)}{w(\xi_w^+)}\frac{a_+}{a_-} \right)^{1/4},
\end{align}
respectively. Further evolving the enthalpy and temperature profiles $w(\xi)$ and $T(\xi)$ backward in $\xi$ into the region $\xi<\xi_{w}$ via Eq.~\eqref{eq:enthalpy_wibag} and~\eqref{eq:temperature} from the above boundary conditions at $(\xi_w^-, w(\xi_w^-))$ and $(\xi_w^-, T(\xi_w^-))$, respectively, one finally obtains their constant values within $\xi<c_{s,-}$. The resulted enthalpy and temperature profiles for the hybrid mode are shown in Fig.~\ref{fig:hybridwt_wibag}.

\section{Bubble expansion beyond the bag equation of state}\label{sec:wobag}

In this section, we will turn to a more general and realistic EoS than a simple bag one. We then propose an iterative method to solve for the fluid EoMs, and finally identify the changes to the efficiency factor of bulk fluid motions.

\subsection{Thermodynamics}\label{subsec:thermodynamics_wobag}

To obtain a more general and realistic EoS than a simple bag one, we start with the effective potential~\cite{Jackiw:1974cv,Dolan:1973qd}, $V_\mathrm{eff}(\phi,T)=V_0(\phi)+V_T(\phi,T)$, consisting of a zero-temperature part $V_0$ and a finite-temperature part up to the one-loop order,
\begin{align}
V_T(\phi,T)= \sum_{i=\mathrm{B},\mathrm{F}}\pm g_i T
    \int \frac{\mathrm{d}^3 \vec{k}}{(2\pi)^3} \ln\left[ 1 \mp e^{-\frac{
    \sqrt{\vec{k}^2 + m^2_i}}{T}} \right]
    \equiv \frac{T^4}{2\pi^2} \sum_{i=\mathrm{B},\mathrm{F}} g_i J_i \left( \frac{m_i^2}{T^2} \right),
\end{align}
where $g_i$ is the number of degrees of freedom for particle species $i$. The integrals for the bosons and fermions,
\begin{align}
    J_{\mathrm{B/F}}(x) = \pm \int_0^\infty \mathrm{d}y~y^2\ln(1\mp e^{-\sqrt{x+y^2}}),
\end{align}
receives exponentially suppressed contributions 
\begin{align}\label{eq:JBF_lowT}
J_\mathrm{B/F}\left(\frac{m_i^2}{T^2}\right)=-\left(\frac{m_i}{2\pi T}\right)^\frac32e^{-m_i/T}\left[1+\mathcal{O}\left(\frac{T}{m_i}\right)\right]
\end{align}
from heavy particles with their masses much larger than the background temperature, $m_i\gg T$. The main contribution to $V_T$ comes from the particle species with $m_i<T$, in which case the integrals can be expanded~\cite{Quiros:1999jp} as
\begin{align}\label{eq:JB}
J_{\mathrm{B}}\left(\frac{m_i^2}{T^2}\right)=
& -\frac{\pi^4}{45}+\frac{\pi^2}{12}\left(\frac{m_i}{T}\right)^2-\frac{\pi}{6}\left(\frac{m_i}{T}\right)^3-\frac{1}{16}\left(\frac{m_i}{T}\right)^4\ln\frac{m_ie^{\gamma_E-3/4}}{4\pi T}\nonumber\\ 
&-\frac{1}{8\pi^{1/2}}\left(\frac{m_i}{T}\right)^4\sum\limits_{l=1}^{\infty}(-1)^l\frac{\zeta(2l+1)}{(l+1)!}\Gamma\left(l+\frac12\right)\left(\frac{m_i^2}{4\pi^2T^2}\right)^l
\end{align}
for bosons and
\begin{align}\label{eq:JF}
J_{\mathrm{F}}\left(\frac{m_i^2}{T^2}\right)=
&-\frac{7}{8}\frac{\pi^4}{45}+\frac{\pi^2}{24}\left(\frac{m_i}{T}\right)^2+\frac{1}{16}\left(\frac{m_i}{T}\right)^4\ln\frac{m_ie^{\gamma_E-3/4}}{\pi T}\nonumber\\ 
&+\frac{1}{4\pi^{1/2}}\left(\frac{m_i}{T}\right)^4\sum\limits_{l=1}^{\infty}(-1)^l\frac{\zeta(2l+1)}{(l+1)!}\left(1-\frac{1}{2^{2l+1}}\right)\Gamma\left(l+\frac12\right)\left(\frac{m_i^2}{\pi^2T^2}\right)^l
\end{align}
for fermions. The relativistic light particles with $m_i\ll T$ simply resemble the perfect thermal gas with a bag EoS, $p_\mathrm{rad}/\rho_\mathrm{rad}=1/3$ from
\begin{align}
V_T^{m_i\ll T}(\phi,T)=-\frac13\frac{\pi^2}{30}\left(\sum\limits_{i=\mathrm{B}}g_i+\frac78\sum\limits_{i=\mathrm{F}}g_i\right)T^4\equiv-\frac13\frac{\pi^2}{30}g_\mathrm{eff}T^4\equiv-\frac13\rho_\mathrm{rad}\equiv-p_\mathrm{rad}.
\end{align}
As for the particle species with their masses comparable to the background temperature, $m_i\lesssim T$, they contribute not only to the leading-order thermal-gas term, but also to the higher-order terms in $m_i/T$. For the case with $m_i<T$, we will truncate it at the linear term in temperature,
\begin{align}\label{eq:VeffbeyondbagEoS}
V_{\mathrm{eff}}(\phi,T)=V_{0}(\phi) -\frac{1}{3}a T^4 + b T^2 - c T,
\end{align}
for our general EoS, where coefficients read
\begin{equation}
\begin{aligned}
a =& \frac{\pi^2}{30} \left(\sum\limits_{i=\mathrm{B}}g_i+\frac78\sum\limits_{i=\mathrm{F}}g_i \right), \\
b =& \frac{1}{24} \left( \sum_{i=\mathrm{B}}g_i m_i^2 + \frac{1}{2}\sum_{i=\mathrm{F}}g_i m_i^2 \right) , \\
c =& \frac{1}{12\pi} \sum_{i=\mathrm{B}}g_i m_i^3. \label{eq:coff}
\end{aligned}
\end{equation}

Four comments are given in order below concerning about the above truncated expansion:

First, we will check in the appendix~\ref{sec:higher_order} that the higher-order contributions, for example, a logarithmic term in temperature, does not significantly change our conclusion.

Second, although the contributions to the effective potential from much heavy particles with $m_i\gg T$ are exponentially suppressed and thus can be safely neglected, the contributions from slightly heavier particles with $m_i\gtrsim T$ does not follow the above truncated expansion and cannot be simply neglected. However, as long as the number of degrees of freedom of these slightly heavy particles with $m_i\gtrsim T$ is not too large, their contributions to the effective potential would be smaller than those truncated terms in the high-temperature expansion of the effective potential. We will discuss this special case in more details in appendix~\ref{sec:Massive_particles}.

Third, for a steadily expanding bubble long after its nucleation, the thin-wall approximation becomes more and more valid so that the scalar field profile in the symmetric and broken phases takes the vacuum expectation values $\phi_\pm$, respectively, thus the dimensional parameters $b$ and $c$ in the above truncated expansion are evaluated with the field-dependent effective masses $m_i$ at the vacuum expectation values $\phi_\pm$ in the symmetric and broken phases, respectively. Since we have assumed no temperature dependence in the vacuum expectation values $\phi_\pm$, the effective masses $m_i$ and the associated parameters $b$ and $c$ in the truncated expansion should be all independent of the temperature.

Finally, for the typical choices of the parameter values in the truncated expansion, we could consider two illustrative examples: one is that the particle contents are the same as the SM like the SMEFT with a dimension-six operator $|H|^6$~\cite{Ellis:2018mja,Postma:2020toi}, where the truncated expansion parameters can be estimated in a dimensionless manner as $a_-\simeq 35$, $b_-/(a_-T^2)\simeq 3\times 10^{-3}$ and $c_-/(a_-T^3)\simeq 2\times 10^{-3}$ 
at a phase transition temperature $T_N\simeq100$ GeV as shown in  appendix~\ref{sec:Massive_particles}; the other one is that there are more particle species than the SM, for example, a PT model at the EW scale with several extra heavy particles with $m_i\sim\mathcal{O}(10^2)$ GeV and a phase transition temperature $T_N\sim 500$ GeV, in which case the dimensionless $b_-/(a_-T_N^2)$ could be estimated as
\begin{align}
    \frac{b_-}{a_-T_N^2}\sim \frac{\sum_i g_i\times(m_i/100\,\mathrm{GeV})^2}{\mathrm{total~number~of~degrees~of~freedom}} \left(\frac{1}{5}\right)^2 \sim \frac{\mathcal{O}(1)}{5^2}.
\end{align}
The same analysis also leads to an estimation for $c_-/(a_-T^3)\sim\mathcal{O}(1)/5^3$. This gives rise to a rough range for the dimensionless parameters $b_-/(a_-T_N^2)\sim\mathcal{O}(10^{-3}-10^{-1})$ and  $c_-/(a_-T^3)\sim\mathcal{O}(10^{-3}-10^{-2})$. Later in this paper, we will adopt a typical set of parameter choice with $T_N=500$ GeV, $b_-/(a_- T_N^2)=2/5^2$ and $c_-/(a_- T_N^3) = 2/5^3$ for illustrations.

With ansatz~\eqref{eq:VeffbeyondbagEoS} for $V_{\mathrm{eff}}$, the pressure and energy density can be directly computed as
\begin{align}
    p &= -\mathcal{F} = -V_{\mathrm{eff}}= -V_{0}(\phi) + \frac{1}{3}a T^4 - b T^2 + cT, \label{eq:pobag}\\
    \rho &= T\partial_T p - p = V_{\mathrm{eff}} - T\partial_T V_{\mathrm{eff}} = V_{0}(\phi) + a T^4 -b T^2 , \label{eq:rhoobag}
\end{align}
respectively, which recovers the bag EoS at the leading order,
\begin{align}
    p_\pm^{(0)} = \frac{1}{3}a_\pm T_\pm^4 - \epsilon_\pm, \quad \rho_\pm^{(0)} = a_\pm T_\pm^4 + \epsilon_\pm,
\end{align}
with abbreviations $\epsilon_\pm\equiv V_0(\phi_\pm)$. The sound velocity also acquires a deviation from $1/3$ by
\begin{align}
    c_s^2 =\frac13\left(1-\frac{4bT-3c}{4aT^3-2bT}\right)= \frac{1}{3} - \frac{b}{3a T^2} + \frac{c}{4a T^3} - \frac{b^2 }{6a^2 T^4} + \mathcal{O}(T^{-5}) . \label{eq:cs2obag}
\end{align}

\subsection{Hydrodynamics}\label{subsec:hydrodynamics_wobag}

With the assumed general EoS for the total scalar-plasma system, we then work out the hydrodynamic solutions and associated energy budget distributions.

\subsubsection{Junction condition}

Although our general EoS~\eqref{eq:pobag} and~\eqref{eq:rhoobag} could still be written as a bag-like form,
\begin{align}
p(T)&=\frac13A(T)T^4-\epsilon(T),\\
\rho(T)&=A(T)T^4+\epsilon(T),
\end{align}
if one defines
\begin{align}
A(T)&=a-\frac{3b}{2T^2}+\frac{3c}{4T^3},\\
\epsilon(T)&=V_0+\frac{b}{2}T^2-\frac{3c}{4}T,
\end{align}
and the junction conditions~\eqref{eq:vbptimesvbm} and~\eqref{eq:vbpdividevbm} also remain unchanged if one defines
\begin{align}\label{eq:alphaplusr}
\alpha_+=\frac{\epsilon_+-\epsilon_-}{A_+T_+^4}\equiv\frac{4 \Delta\epsilon}{3w_+}, \quad r=\frac{A_+T_+^4}{A_-T_-^4}\equiv\frac{w_+}{w_-},
\end{align}
the resulted fluid profiles of velocity, enthalpy, and temperature would explicitly depend on the ratio $A_+/A_-$, which cannot be known \textit{a priori} from a given particle physics model due to its explicit temperature dependence that is only known after solving the fluid EoMs. In order for the physical parameter $a_+/a_-$ from a given particle physics model as the input for hydrodynamic evaluations, we adopt the definitions for
\begin{align}
\alpha_+=\frac{\epsilon_+-\epsilon_-}{a_+T_+^4}, \quad r=\frac{a_+ T_+^4}{a_-T_-^4}, 
\end{align}
with $\epsilon_\pm\equiv V_0(\phi_\pm)$. Furthermore, since particles in front of the wall are in symmetric phase without masses, we can then regard that the bag EoS is still valid outside the wall by assuming $b_+ = c_+ = 0$. Nevertheless, the broken phase still admits our general EoS beyond the bag model. Therefore, the local wall-frame fluid velocities can be solved similarly as
\begin{align}
\bar{v}_+&=\sqrt{\frac{F(T_-)-(1-3\alpha_+)r}{G(T_-)-3(1+\alpha_+)r}\cdot
\frac{G(T_-)+(1-3\alpha_+)r}{F(T_-)+3(1+\alpha_+)r}}, \label{eq:vbpobag} \\
\bar{v}_-&=\sqrt{\left.\frac{F(T_-)-(1-3\alpha_+)r}{G(T_-)-3(1+\alpha_+)r}
\right/\frac{G(T_-)+(1-3\alpha_+)r}{F(T_-)+3(1+\alpha_+)r}}, \label{eq:vbmobag}
\end{align}
with $T_\pm$ to be the temperature just in the front and back of the bubble wall, and
\begin{align}
F(T) &= 1- \frac{3b_-}{a_-} \frac{1}{T^2} + \frac{3c_-}{a_-} \frac{1}{T^3} , \label{eq:FT} \\
G(T) &= 3- \frac{3b_-}{a_-} \frac{1}{T^2} .  \label{eq:GT}
\end{align}

\subsubsection{Hydrodynamic regime}

The hydrodynamic regime parametrized by $\bar{v}_+$ as a function of $\bar{v}_-$ after eliminating $r$ is different from the bag case due to the small deviations from $1$ and $3$ in $F(T)$ and $G(T)$, respectively. To gain more physical insight from analytical estimations, we will take $c_-=0$ for the moment as an illustration, but the physical picture also holds numerically for $c_-\neq0$.

\begin{figure}[t]
    \centering
    \subfigure[$\alpha_+>\alpha_c$]{
    \includegraphics[width=0.3\textwidth]{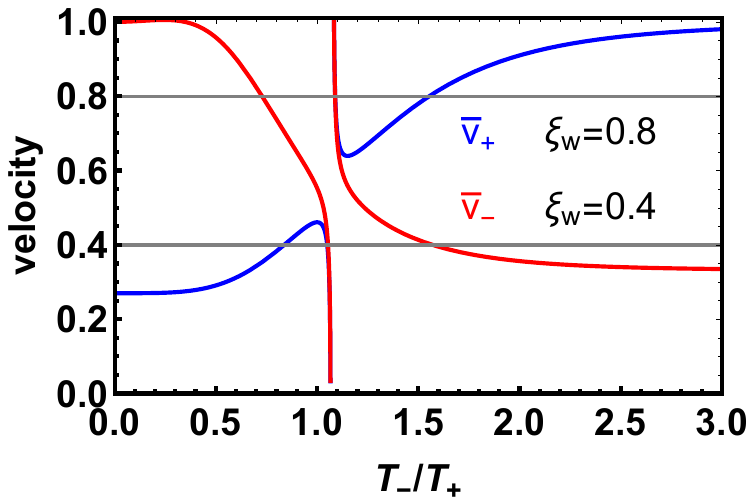}
    \label{fig:gtac}
    }
    \subfigure[$\alpha_+=\alpha_c$]{
    \includegraphics[width=0.3\textwidth]{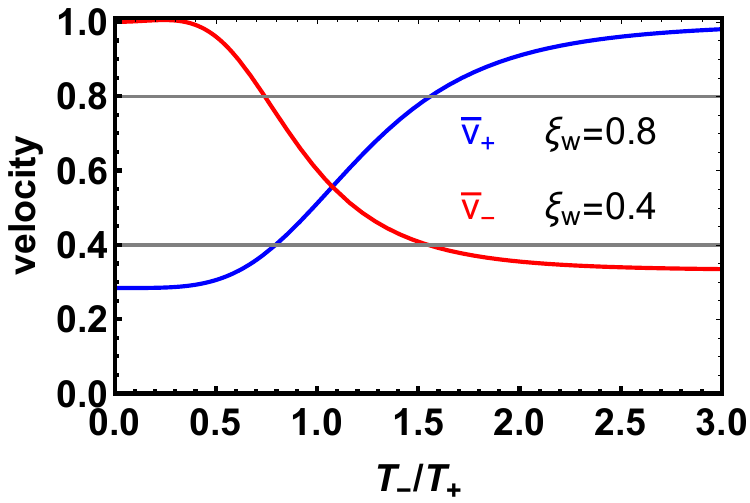}
    \label{fig:etac}
    }
    \subfigure[$\alpha_+<\alpha_c$]{
    \includegraphics[width=0.3\textwidth]{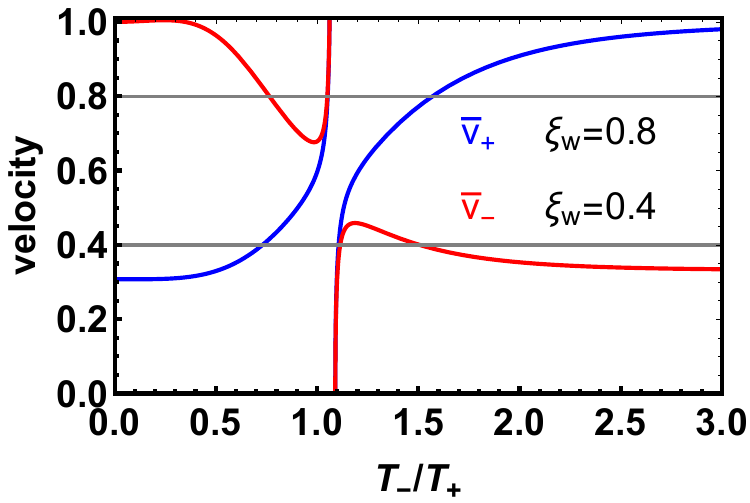}
    \label{fig:ltac}
    }
    \caption{$\bar{v}_\pm$ as functions of $t\equiv T_-/T_+$ for a typical set of the parameter choices with $K =a_+/a_-=1.2$, $B=0.08$, and (a) $\alpha_+=0.04$; (b) $\alpha_+=\alpha_c\approx 0.03$; (c) $\alpha_+=0.02$. The blue and red curves correspond to $\bar{v}_+$ and $\bar{v}_-$, respectively, while the two horizontal gray lines denote $v=\xi_w=0.8$ and $v=\xi_w=0.4$. When the line $v=\xi_w$ is tangent to the right branch of the $\bar{v}_+$ curve at its minimum in situation (a), one obtains the Jouguet detonation and the corresponding Jouguet velocity at the minimum of the right branch of $\bar{v}_+$  curve.}
    \label{fig:vpvmofT}
\end{figure}

First, unlike the bag EoS case with $\alpha_+>0$, there is a positive lower bound on $\alpha_+$ beyond the bag EoS. This can be most easily seen from the illustrative case below with $c_-=0$ where the lower bound on $\alpha_+$ can be worked out analytically. After abbreviating 
\begin{align}
t\equiv \frac{T_-}{T_+}, \quad B = \frac{b_-}{a_- T_+^2}, \quad K = \frac{a_+}{a_-},
\end{align}
Eq.~\eqref{eq:vbpibag} now reads
\begin{align}
    \bar{v}_+ = \sqrt{\frac{1-3B t^{-2}-(1-3\alpha_+)Kt^{-4}}{3-3Bt^{-2}-3(1+\alpha_+)Kt^{-4}} \cdot \frac{3-3B t^{-2}+(1-3\alpha_+)Kt^{-4}}{1-3Bt^{-2}+3(1+\alpha_+)Kt^{-4}}}.
\end{align}
The second factor in the square root comes from $(\rho_- + p_+)/(\rho_+ + p_-)$, which must be greater than $0$, hence the first factor has to be positive too. The same analysis also holds for $\bar{v}_-$. Therefore, the local wall-frame fluid velocities $\bar{v}_+$ and $\bar{v}_-$ as functions of $T_-/T_+$ can be presented in Fig.~\ref{fig:gtac} for a not-too-small $\alpha$, in which case $\bar{v}_\pm$ are split into two branches corresponding to the detonation and deflagration modes as we will identify later. When $\alpha$ is decreasing, the two branches become closer and closer until finally join together into one curve as shown in Fig.~\ref{fig:etac}. This happens at $\alpha_+=0$ from Eq.~\eqref{eq:vbpibag} and~\eqref{eq:vbmibag} for the bag EoS, but a finite value for $\alpha_+$ when we go beyond the bag EoS from Eq.~\eqref{eq:vbpobag} and~\eqref{eq:vbmobag}. We denote this critical lower bound of $\alpha_+$ as $\alpha_c$. Further decreasing $\alpha_+$ to cross this critical point, the two branches simply swap their position as shown in Fig.~\ref{fig:ltac}, which is physically forbidden similar to the mathematical continuation analog of $\alpha<0$ in the bag model. To be specific, the lower bound on $\alpha_+$ comes from the requirement of keeping $\bar{v}_+ \bar{v}_- = (p_+ - p_-)/(\rho_+ - \rho_-) >0$, which would impose an extra constraint on the our EoS parameters $a$, $b$, and $c$. If a particle physics model of a FOPT fails to meet the lower bound on $\alpha_+$, then there would be a difficulty of plugging the junction conditions into fluid EoM. Note that this does not imply anything wrong about such a particle physics model but the limitation of our EoS ansatz in describing this particle physics model.

\begin{figure}
    \centering
    \subfigure[$\bar{v}_+$ of $\bar{v}_-$ for various $\alpha_+$]{
    \includegraphics[width=0.47\textwidth]{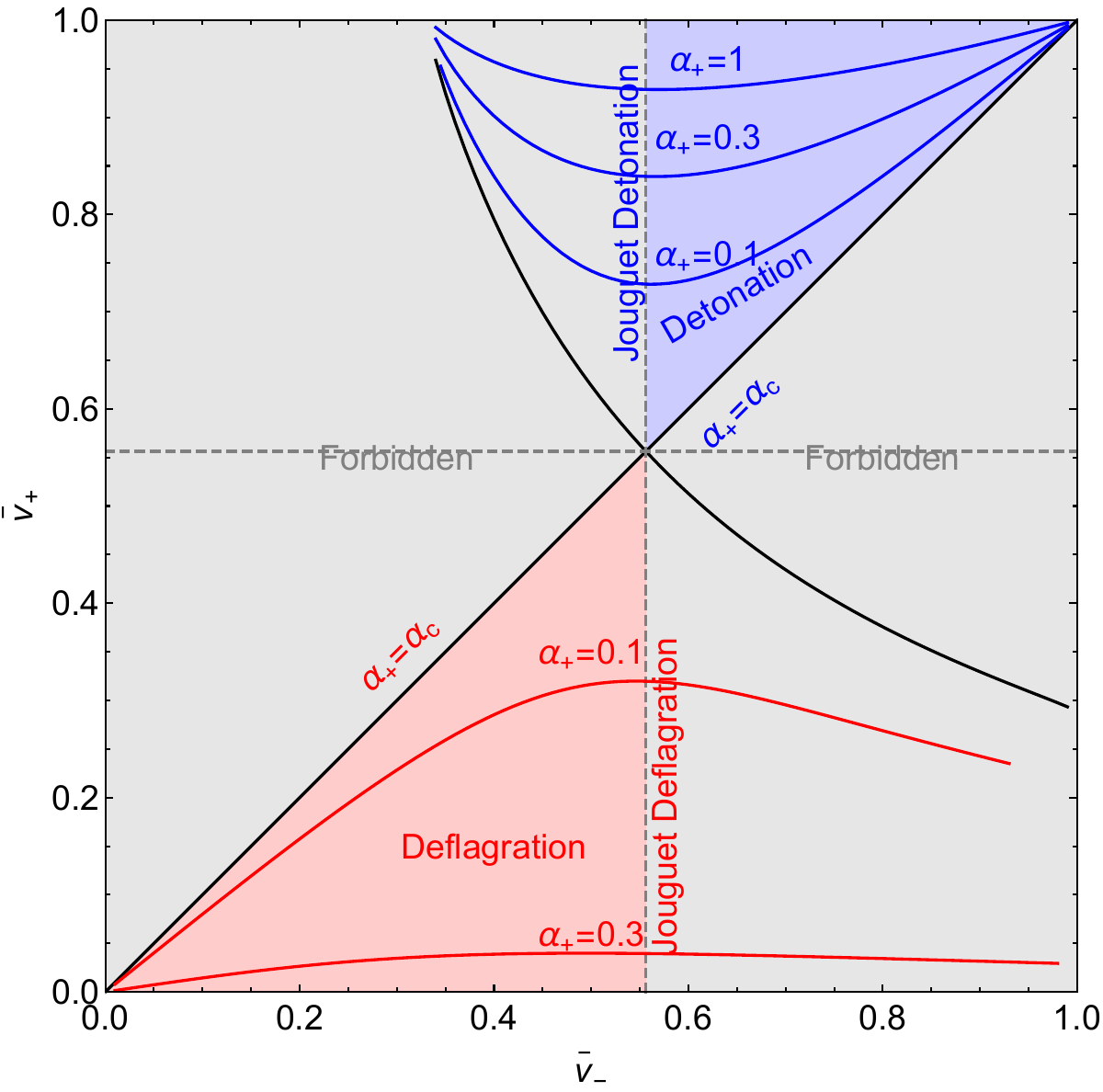}
    \label{fig:vmvpofalpha}
    }
    \subfigure[Jouguet velocity]{
    \includegraphics[width=0.47\textwidth]{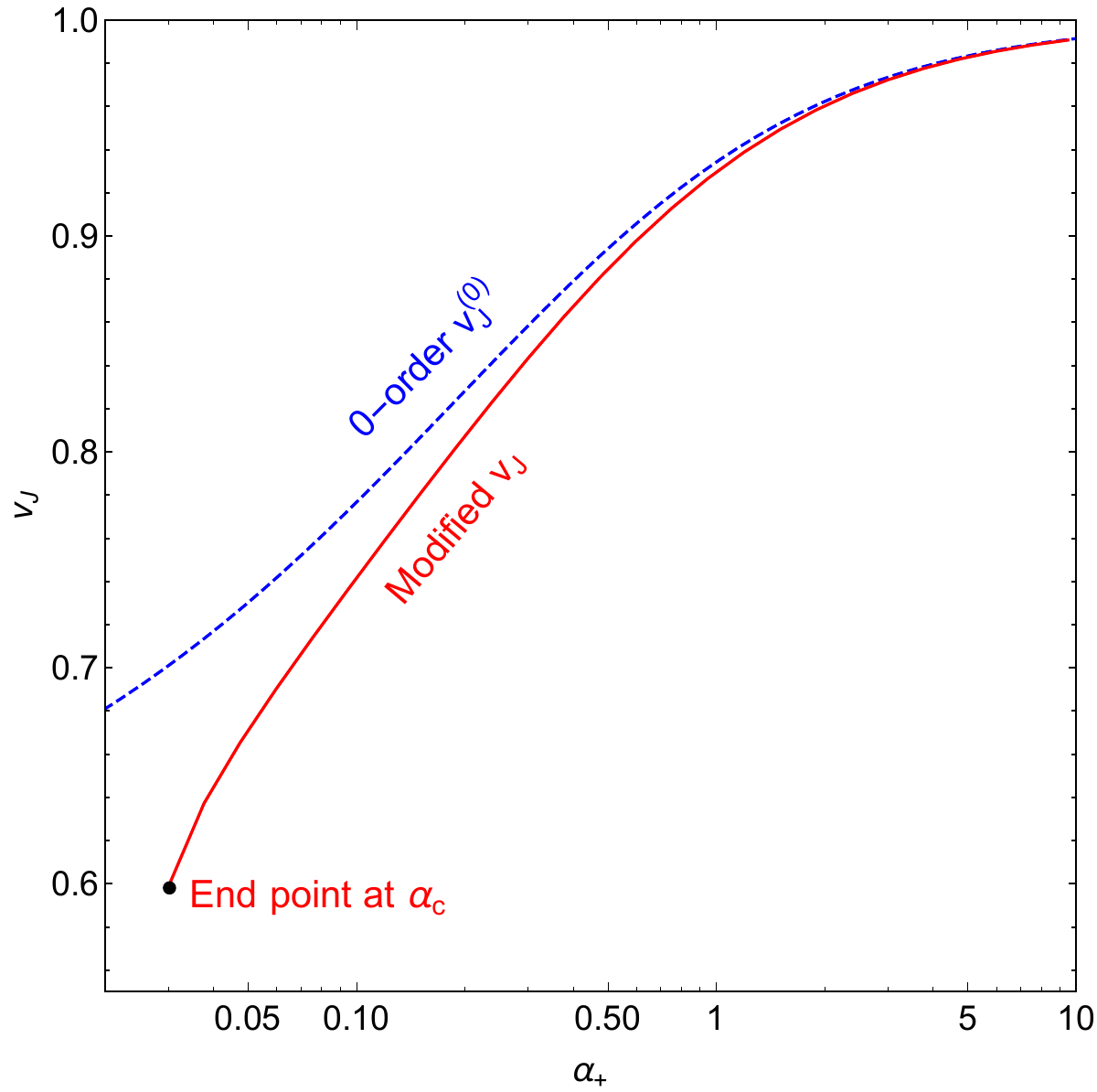}
    \label{fig:vJ}
    }
    \caption{(a) The hydrodynamic regime determined by $\bar{v}_+$ as a function of $\bar{v}_-$ for fixed $\alpha_+$. The typical parameters choices are the same as Fig.~\ref{fig:vpvmofT}. The horizon and vertical gray dashed lines are located at $\bar{v}_c$ determined by Eq.~\eqref{eq:vc}.
    (b) The Jouguet velocity with (red solid) and without (blue dashed) the modifications from $b_-/(a_- T_N^2)=2/25$, $c_-/(a_- T_N^3)=2/125$ and $T_N = 500$ GeV. The endpoint is at $(v_\mathrm{J},\alpha_c)\approx(0.60,0.025)$.}
\end{figure}

Next, to determine the value of $\alpha_c$, we first note that the discontinuity point $t=t_0$ between two branches only disappears when $\alpha_+ = \alpha_c$, thus we can expect that $t_0$ in this case should be the root of both the numerator and denominator of the first factor in the square root,
\begin{align}
    1 - 3\alpha_c =& K^{-1}t_0^4( 1-3B t_0^{-2} ), \\
    1 + \alpha_c =& K^{-1}t_0^4( 1-B t_0^{-2} ),
\end{align}
where the positive critical value $\alpha_c$ can be directly solved as
\begin{align}
    \alpha_c = \frac{B}{8K}\left( 3B+\sqrt{9B^2 + 16K^2 } \right). \label{eq:alphac}
\end{align}
For $\alpha_+=\alpha_c$, the local wall-frame velocity at the joining point deviates from $\bar{v}_c = 1/\sqrt{3}$ by 
\begin{align}
    \bar{v}_c^2 = \frac13\left(1-\frac{6B}{3B+\sqrt{16K+9B^2}}\right),  \label{eq:vc}
\end{align}
which can be shown as the horizontal and vertical gray dashed lines in Fig.~\ref{fig:vmvpofalpha}. Similarly, the upper limit for $\alpha_+$ in deflagration region also slightly deviates from $1/3$. It is also worth noting that, $\bar{v}_+$ as a function of $\bar{v}_-$ in Fig.~\ref{fig:vmvpofalpha} for the deflagration (red curves) and detonation (blue curves) modes admits an endpoint in the physically forbidden regions (gray shaded). For example, the deflagration curve with $\alpha_+=0.1$ has an endpoint not at $\bar{v}_-=1$ but at $\bar{v}_-\approx 0.93$. This can be traced back to the moment when solving Eq.~\eqref{eq:vbpobag} and~\eqref{eq:vbmobag}, one must always keep $T_-^2>0$ that leads to a constraint on the $\bar{v}_--\bar{v}_+$ plane. One can also find similar endpoints at the very left of detonation curves. Fortunately, these endpoints do not bother us since they are located at the physically forbidden ``strong detonation'' and ``strong deflagration'' regions. All the discussions above also hold for $c_-\neq0$ but with different numerics. 

Last but not least, the Jouguet velocity is also modified compared to the case~\eqref{eq:vJibag} with a bag EoS. Recall that the Jouguet velocity corresponds to the point when the horizontal line $v=\xi_w$ is tangent to the curve $\bar{v}_+$ as a function of $T_-/T_+$ at its local minimum point for a given $\alpha_+$, where $T_+ = T_N$ if we approach the Jouguet velocity from the detonation regime. For a general $c_-\neq0$, the modified Jouguet velocity can be solved numerically as shown in Fig.~\ref{fig:vJ}, which is lower than the case with a bag EoS, but the deviation becomes smaller for a larger $\alpha_+$. Furthermore, since the Jouguet velocity is a function of $\alpha_+$, it also admits an endpoint on the left inherited from the lower bound on $\alpha_+$. The suppression of the Jouguet velocity beyond a bag EoS is consistent with the decrease of the sound velocity in the broken phase as we will see later when we actually solve for the fluid EoMs. This also indicates a suppression in the efficiency factor of fluid motions as we will see in the end, though the suppression is less pronounced for a larger $\alpha_+$.

\subsubsection{Fluid equations of motions}

We next turn to the modifications to the fluid EoM~\eqref{eq:EoMibag} and enthalpy equation~\eqref{eq:enthalpyibag}, which, after written with the similarity coordinate, are derived with similar forms as the bag case,
\begin{align}
2\frac{v}{\xi}&=\gamma(v)^2(1-\xi v)\left(\frac{\mu(\xi,v)^2}{c_s^2(T)}-1\right)\frac{\mathrm{d}v}{\mathrm{d}\xi}, \label{eq:EoMobag}\\
\frac{\mathrm{d}w}{\mathrm{d}\xi}&=w\gamma(v)^2\mu(\xi,v)\left(\frac{1}{c_s^2(T)}+1\right)\frac{\mathrm{d}v}{\mathrm{d}\xi},\label{eq:enthalpyobag}
\end{align}
where the sound velocity $c_s^2(T)=\mathrm{d}p/\mathrm{d}\rho$ evaluated by Eq.~\eqref{eq:cs2obag} is now a spatially dependent function of $\xi$ through its explicitly temperature dependence $T(\xi)$ given by
\begin{align}
T(\xi)=T(\xi_0)\exp\left[\int_{v(\xi_0)}^{v(\xi)}\gamma(v)^2\mu(\xi,v)\mathrm{d}v\right].
\end{align}
Therefore, the modified fluid EoM and enthalpy equation are in fact integro-differential equations, which are difficult to solved directly. In this paper, we propose to solve above equations with perturbative iterations. The algorithm goes as follows:

\begin{enumerate}
\item We start with the bag EoS with $c_s^2=1/3$ to solve for the zeroth-order profiles $v^{(0)}(\xi)$, $w^{(0)}(\xi)$ and $T^{(0)}(\xi)$, which has been done in Sec.~\ref{sec:wibag}.
\item We plug the zeroth-order temperature profile $T^{(0)}(\xi)$ into Eq.~\eqref{eq:cs2obag} to get the first-order sound velocity profile ${c_s^2}^{(1)}(\xi)$, and then solve Eq.~\eqref{eq:EoMobag} for the first-order velocity profile $v^{(1)}(\xi)$ and associated profiles $w^{(1)}(\xi)$ and $T^{(1)}(\xi)$.
\item We repeat the step 2 by plugging $i$-th order temperature profile $T^{(i)}(\xi)$ into Eq.~\eqref{eq:cs2obag} to get the $(i+1)$-order sound velocity profile ${c_s^2}^{(i+1)}(\xi)$ and then solving for the $(i+1)$-order profiles $v^{(i+1)}(\xi)$, $w^{(i+1)}(\xi)$ and $T^{(i+1)}(\xi)$ until all these profiles deviate negligibly from the $i$-th order profiles.
\end{enumerate}
We will see that these profiles converge so fast that there is almost no significant difference between first-order and second-order profiles of the fluid velocity, enthalpy, and temperature. 

\begin{figure}
    \centering
    \subfigure[$v(\xi)$ for detonation]{
    \includegraphics[width=0.47\textwidth]{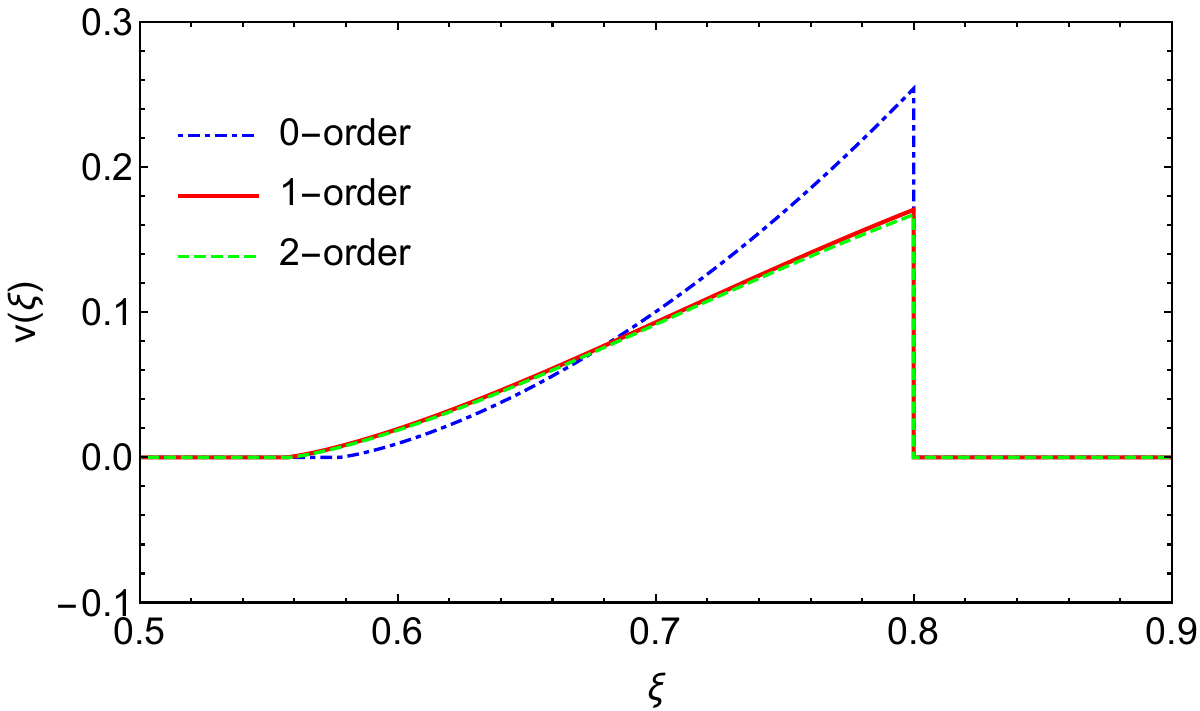}
    \label{fig:detonav_wobag}
    }
    \subfigure[$c_s^2(\xi)$ for detonation]{
    \includegraphics[width=0.47\textwidth]{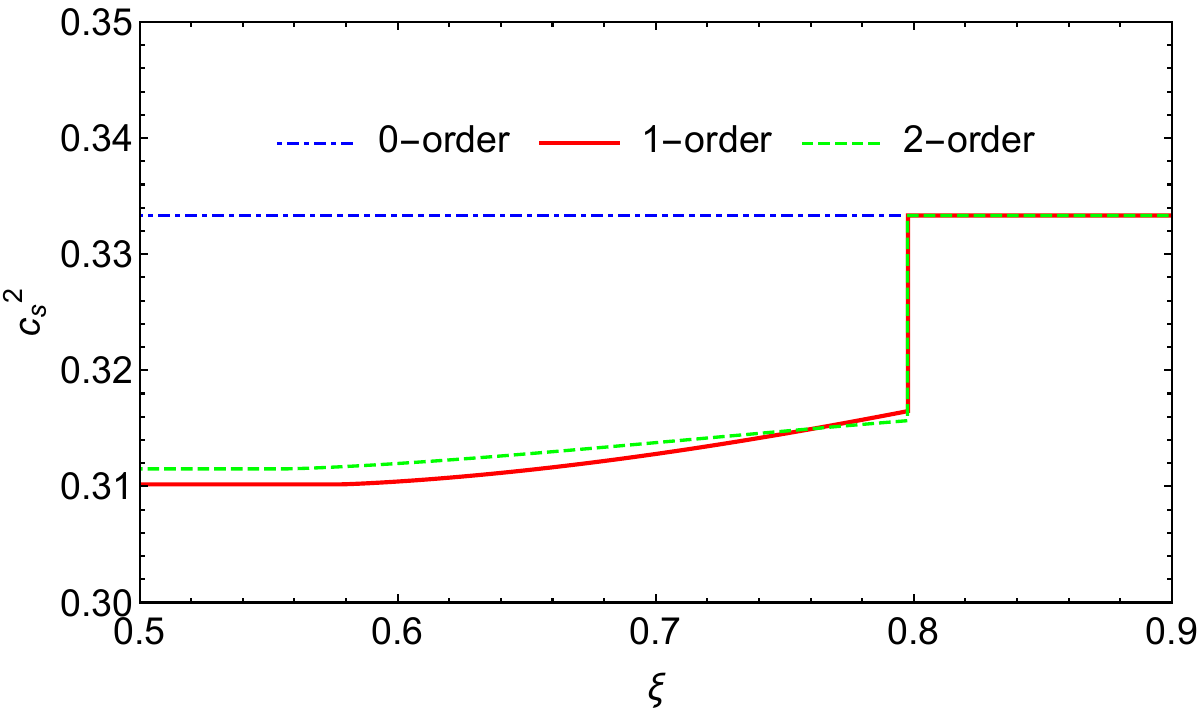}
    \label{fig:detonacs_wobag}
    }
    \subfigure[$v(\xi)$ for hybrid]{
    \includegraphics[width=0.47\textwidth]{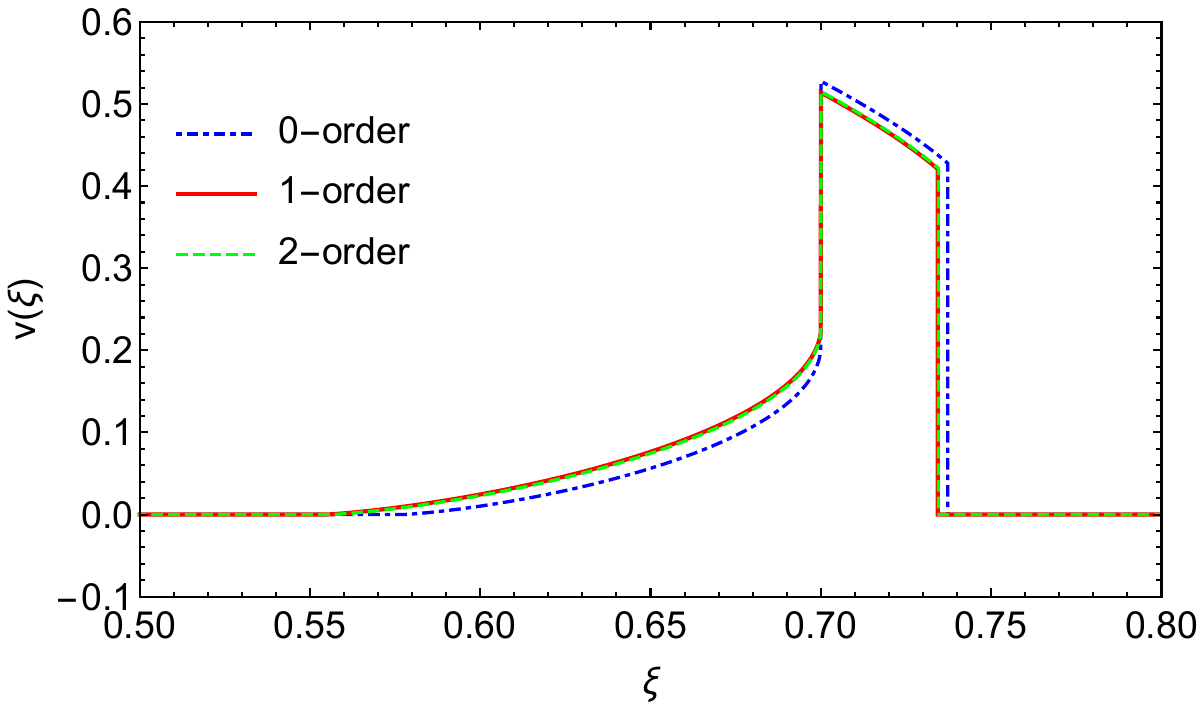}
    \label{fig:hybridv_wobag}
    }
    \subfigure[$c_s^2(\xi)$ for hybrid]{
    \includegraphics[width=0.47\textwidth]{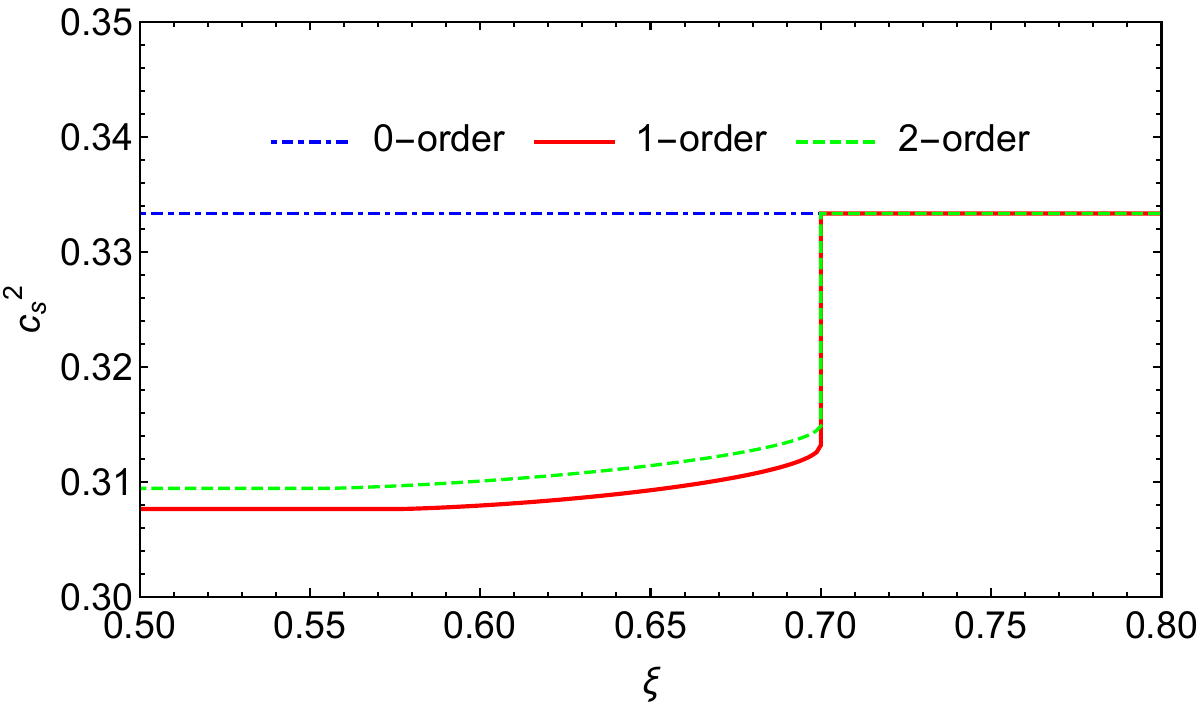}
    \label{fig:hybridcs_wobag}
    }
    \subfigure[$v(\xi)$ for deflagration]{
    \includegraphics[width=0.47\textwidth]{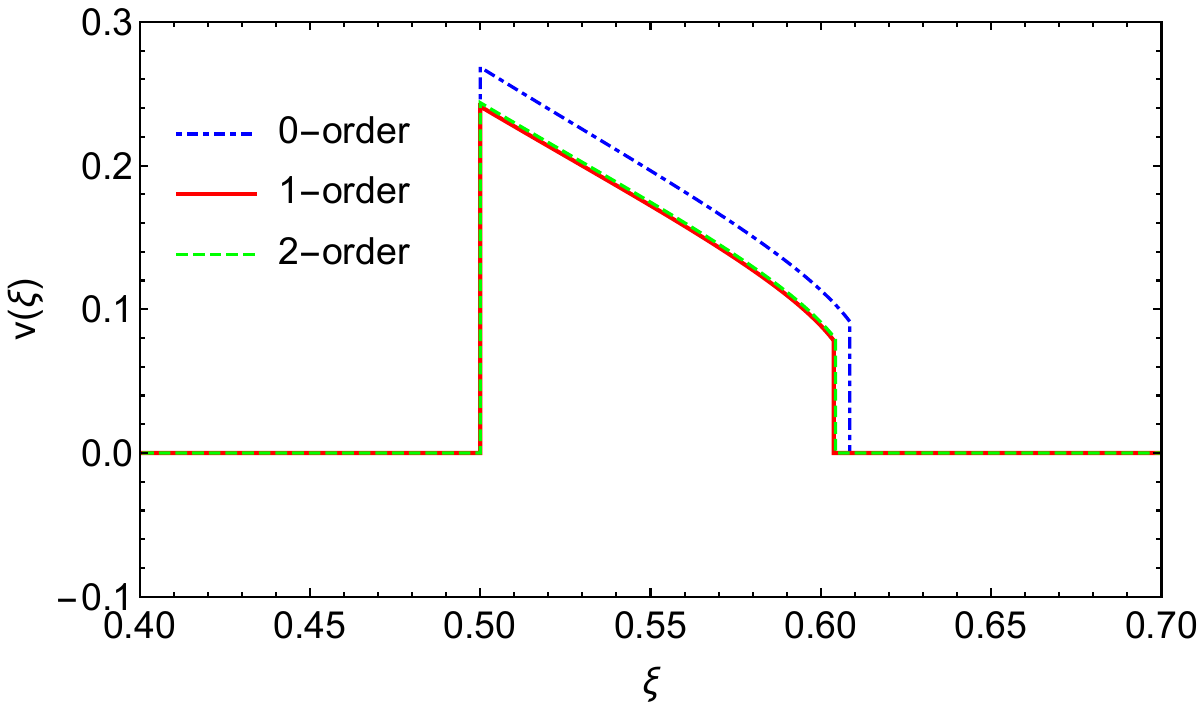}
    \label{fig:deflagv_wobag}
    }
    \subfigure[$c_s^2(\xi)$ for deflagration]{
    \includegraphics[width=0.47\textwidth]{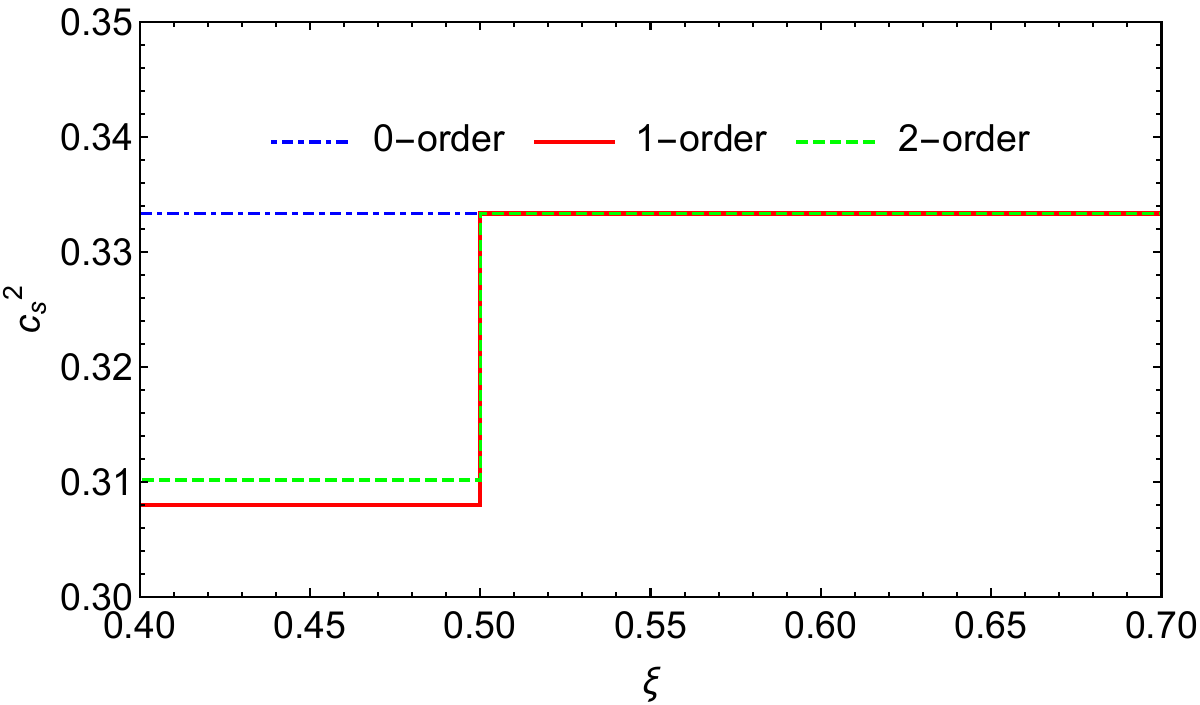}
    \label{fig:deflagcs_wobag}
    }
    \caption{The profiles of the fluid velocity (left column) and sound velocity (right column) for the bubble expansion modes of detonation (top row: $\xi_w=0.8$), hybrid (middle row: $\xi_w=0.7$), and deflagration (bottom row: $\xi_w=0.5$) types are illustrated with a typical set of parameter choices $\alpha_+=0.1$, $a_+/a_-=1.2$, $b_-/(a_- T_N^2)=2/25$, $c_-/(a_- T_N^3)=2/125$, and $T_N = 500$ GeV, where the numerical results from the zeroth-order, first-order, and second-order iterations  are shown with the blue dot-dashed, red solid, and green dahsed curves, respectively. The zeroth-order iteration results are simply from the bag EoS, which is also adopted in the symmetric phase for all cases.}
    \label{fig:velocity_profiles_obag}
\end{figure}

\begin{figure}
    \centering
    \subfigure[$w(\xi)$ for detonation]{
    \includegraphics[width=0.47\textwidth]{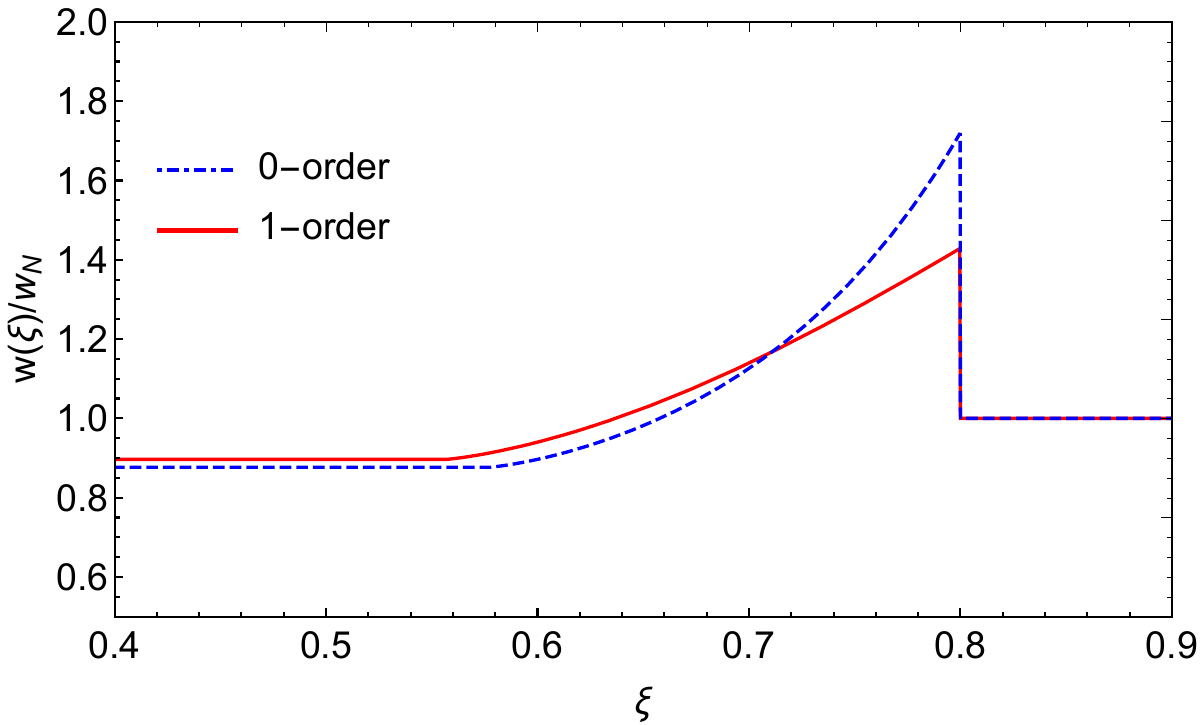}
    \label{fig:detonaw_wobag}
    }
    \subfigure[$T(\xi)$ for detonation]{
    \includegraphics[width=0.47\textwidth]{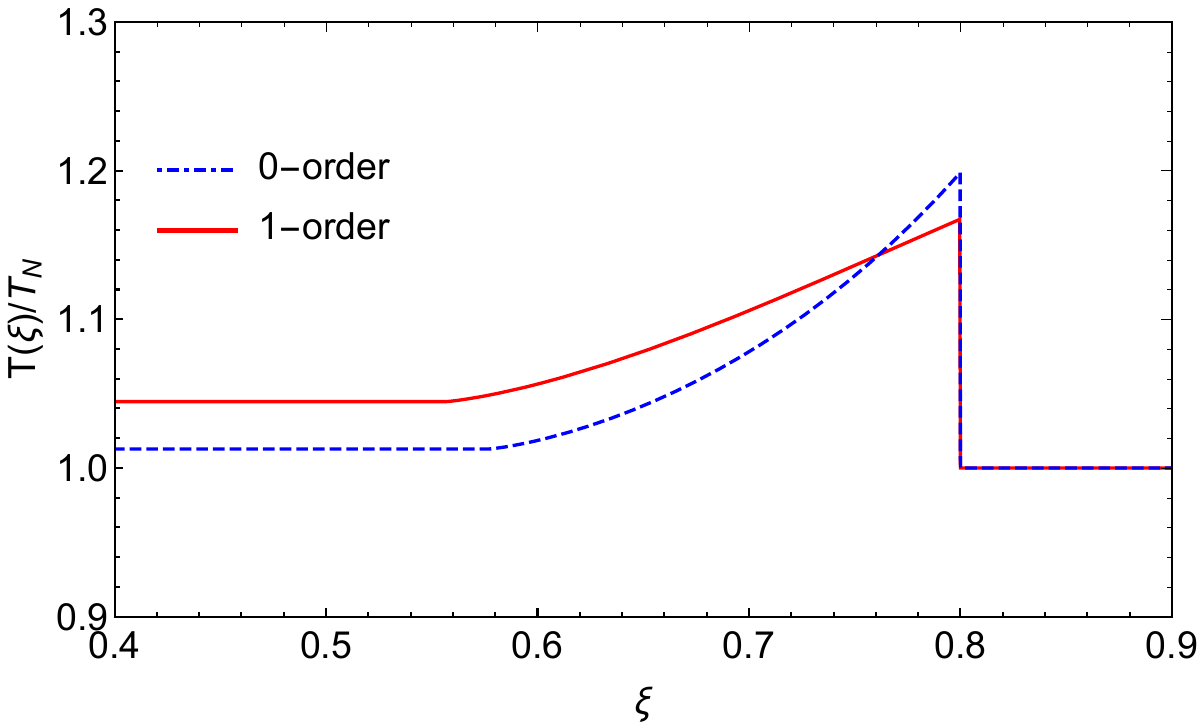}
    \label{fig:detonaT_wobag}
    }
    \subfigure[$w(\xi)$ for hybrid]{
    \includegraphics[width=0.47\textwidth]{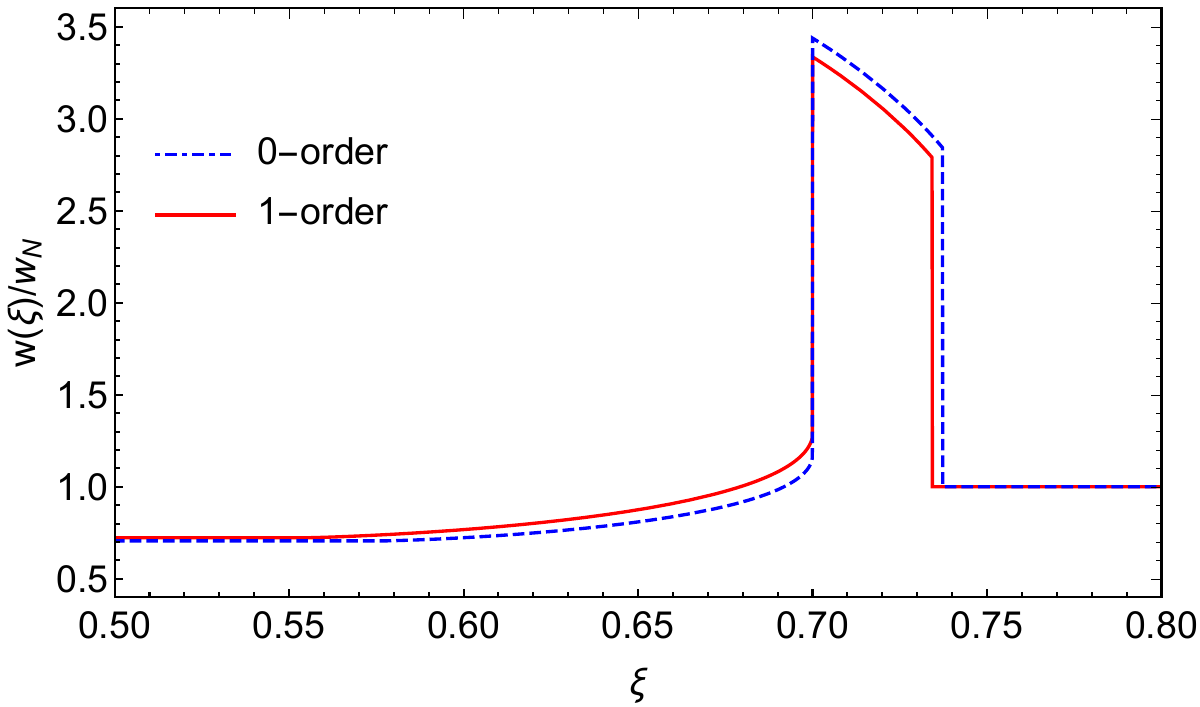}
    \label{fig:hybridw_wobag}
    }
    \subfigure[$T(\xi)$ for hybrid]{
    \includegraphics[width=0.47\textwidth]{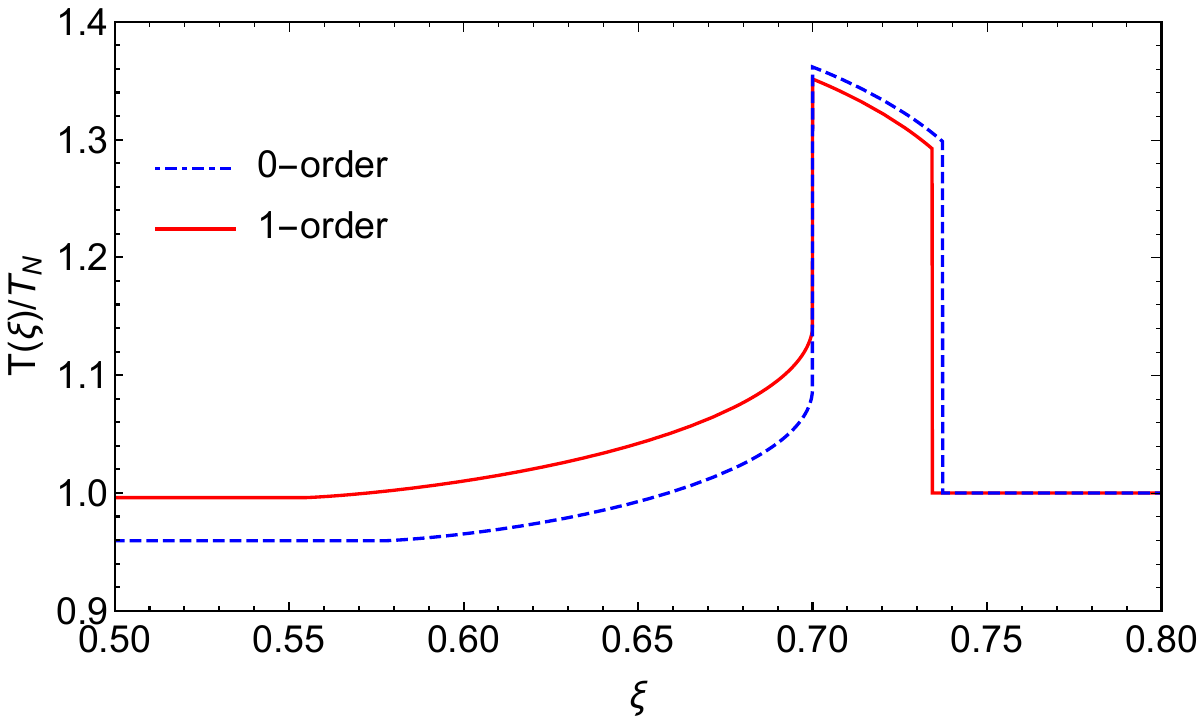}
    \label{fig:hybridT_wobag}
    }
    \subfigure[$w(\xi)$ for deflagration]{
    \includegraphics[width=0.47\textwidth]{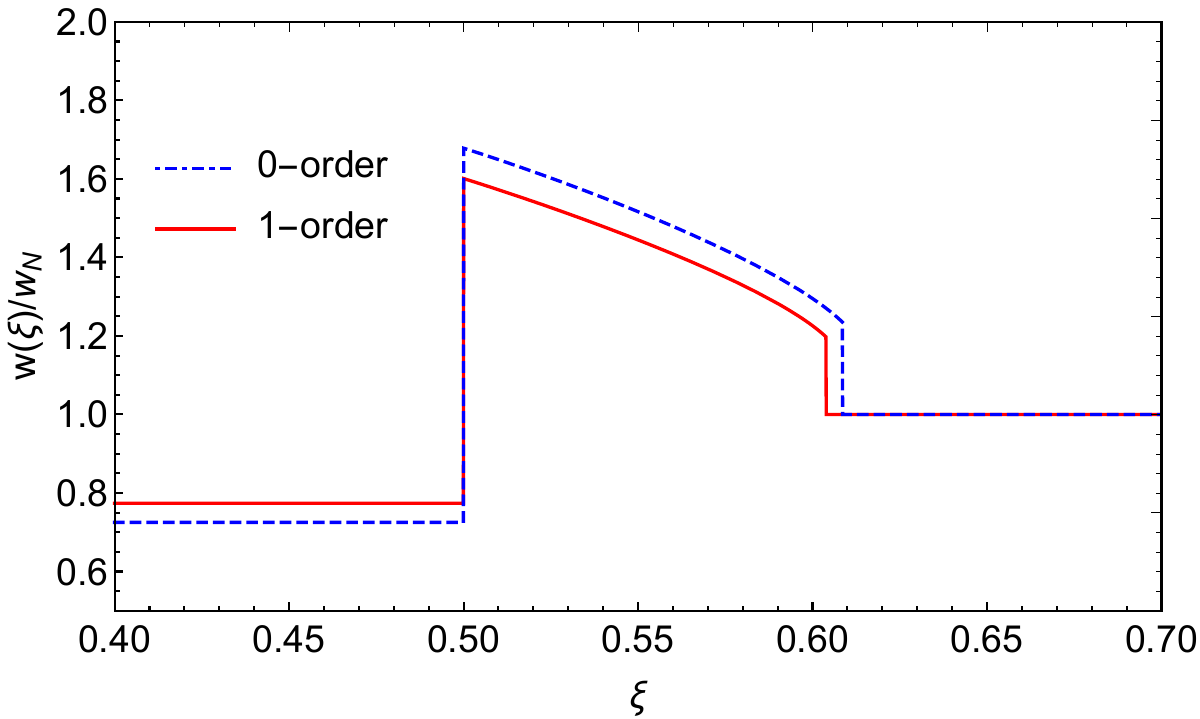}
    \label{fig:deflagw_wobag}
    }
    \subfigure[$T(\xi)$ for deflagration]{
    \includegraphics[width=0.47\textwidth]{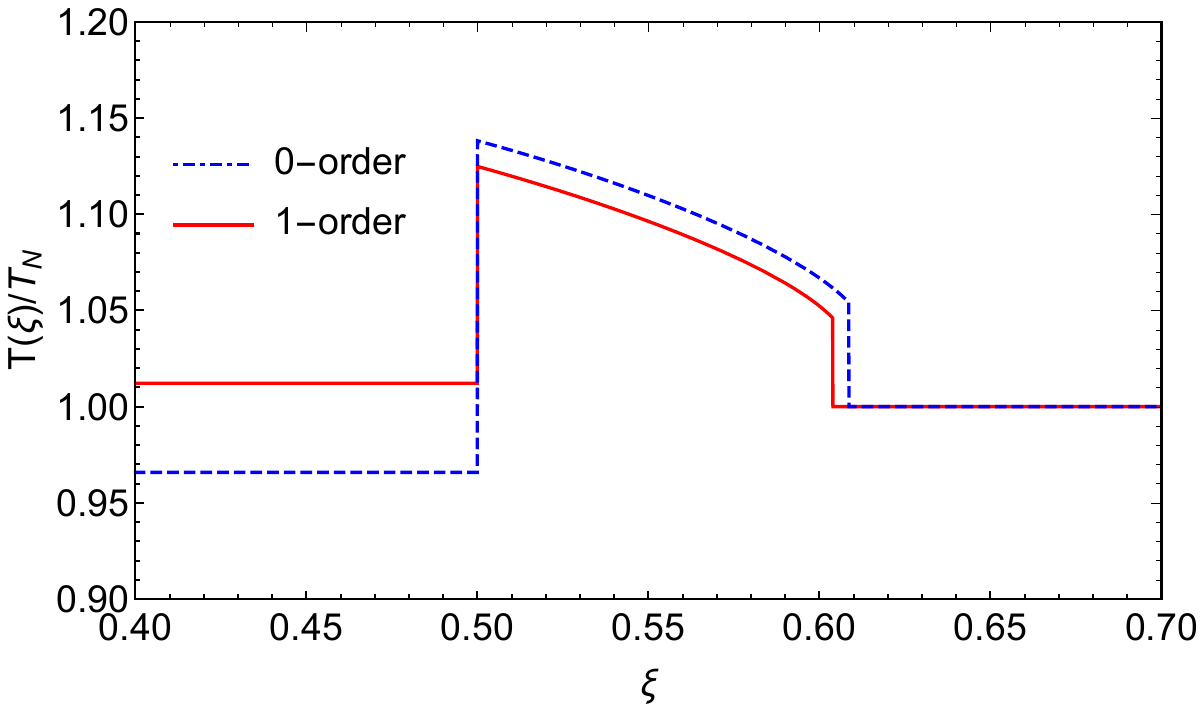}
    \label{fig:deflagT_wobag}
    }
    \caption{The profiles of the enthalpy (left column) and temperature (right column) for the bubble expansion modes of detonation (top row: $\xi_w=0.8$), hybrid (middle row: $\xi_w=0.7$), and deflagration (bottom row: $\xi_w=0.5$) types are illustrated with a typical set of parameter choices $\alpha_+=0.1$, $a_+/a_-=1.2$, $b_-/(a_- T_N^2)=2/25$, $c_-/(a_- T_N^3)=2/125$, and $T_N = 500$ GeV, where the numerical results from the zeroth-order and first-order iterations are shown with the blue dashed and red solid curves, respectively. The bag EoS is adopted in the symmetric phase for all cases.}
    \label{fig:thermal_profiles_obag}
\end{figure}

\subsection{Expansion modes} \label{subsec:expansion_wobag}

In this subsection, we solve the modified fluid EoM~\eqref{eq:EoMobag} with perturbative iterations for the modified profiles of the fluid velocity, enthalpy and temperature. Some illustrative examples of numerical results for the sound velocity and fluid velocity profiles from three types of expansion modes are shown in Fig.~\ref{fig:velocity_profiles_obag} and the corresponding enthalpy and temperature profiles are shown in Fig.~\ref{fig:thermal_profiles_obag}, where the bag EoS case is compared as the zeroth-order input for our iteration method in solving the fluid EoMs with an inhomogeneous profile of the sound velocity. The comparison to the $\nu$-model is also discussed in the appendix~\ref{sec:compare_with_nu}. The efficiency factor of energy budget obtained from these modified results is evaluated in next subsection. 

\subsubsection{Detonation}\label{subsec:detona_wobag}

The detonation of the weak type is defined by $c_s<\bar{v}_-<\bar{v}_+=\xi_w$, which requires the bubble wall velocity $\xi_w>v_J$ larger than the modified Jouguet velocity that is smaller than the original Jouguet velocity~\eqref{eq:vJibag} of the bag case. The boundary condition $\bar{v}_+=\xi_w$ is still derived from recognizing the fluid velocity in front of the bubble wall to be static, namely, $v_+ = 0 = \mu(\xi_w, \bar{v}_+)$. The absence of the sound shell in front of the bubble wall would render $T_+ = T_N$, thus we can use Eq.~\eqref{eq:vbpobag} to inversely express $T_-(\xi_w,\alpha_+)$, from which $\bar{v}_-$ and $v_-=\mu(\xi_w,\bar{v}_-)$ can also be evaluated accordingly. Once the boundary condition for the fluid velocity in the vicinity of the bubble wall is prepared, our perturbative iteration method can be applied immediately to yield a good approximation for the solution of the fluid EoM. An example of such iteration solution is presented in Fig.~\ref{fig:detonav_wobag} along with the corresponding sound velocity shown in Fig.~\ref{fig:detonacs_wobag}. 

\begin{figure}
    \centering
    \includegraphics[width = 4in]{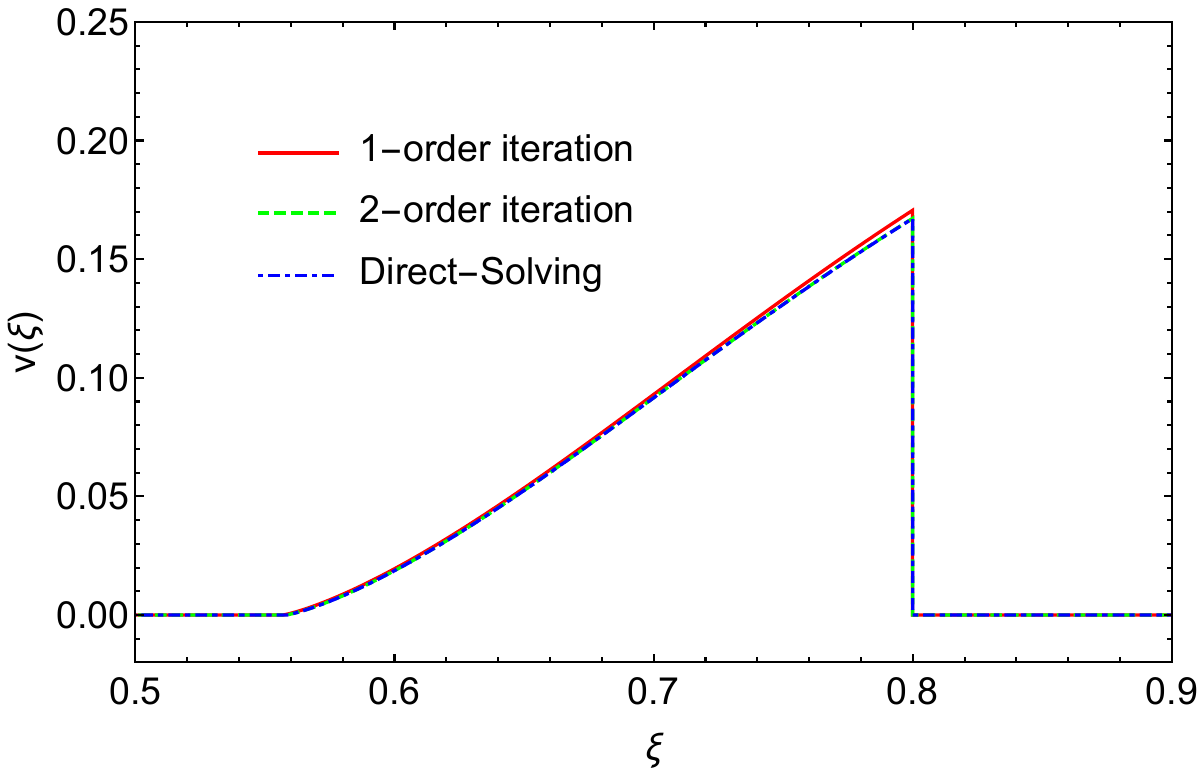}
    \caption{The comparison of the numerical results for the detonation mode from the iteration method and direct-solving method with a typical set of parameter choices $\alpha_+=0.1$, $a_+/a_-=1.2$, and $\xi_w=0.8$.}
    \label{fig:detona_compare}
\end{figure}

However, there is a subtly for the case when the bubble wall velocity lies between the modified Jouguet velocity $v_{\mathrm{J}}$ and the original one $v_{\mathrm{J}}^{(0)}$ from the bag model, in which case our perturbative iteration method would breaks down since there is simply no zeroth-order detonation solution for $\xi_w<v_{\mathrm{J}}^{(0)}$. Therefore, we have to directly solve the integral-differential Eq.~\eqref{eq:EoMobag} for this special case. Nevertheless, for $\xi_w>v_{\mathrm{J}}^{(0)}$, the iteration method is still viable and the velocity profile converges very quickly. In Fig.~\ref{fig:detona_compare}, we have shown the indistinguishable solutions from solving the integral-differential Eq.~\eqref{eq:EoMobag} directly or iteratively for a $\xi_w$ not too close to $v_{\mathrm{J}}^{(0)}$, where the difference between the second-order iteration and direct-solving results is of the order $10^{-5}$. This also validates our perturbative iteration method as a good approximation to the direct solving result.

Once we have solved the profile of the fluid velocity for the detonation mode, we can then follow what we have done in the last section to give rise to the profiles of enthalpy and temperature as shown in Fig.~\ref{fig:detonaw_wobag} and~\ref{fig:detonaT_wobag}. Applying the junction condition across the bubble wall, 
\begin{align}
    w(\xi_w^-) = w_N \frac{\xi_w}{1-\xi_w^2} \frac{ 1 - \bar{v}_-^2}{\bar{v}_-},
\end{align}
with $\bar{v}_-=\bar{v}_-(\alpha_+,T_-(\xi_w,\alpha_+))$, we can obtain the enthalpy profile by evolving $w(\xi)$ from the wall position for $\xi<\xi_w$ as
\begin{align}
     w(\xi) = w(\xi_w^-) \exp\left[ -\int_{v(\xi)}^{v(\xi_w)} \left( 1+\frac{1}{c_s^2(T(\xi(v)))} \right) \gamma^2(v) \mu(\xi(v),v) \mathrm{d}v \right].
\end{align}
Note that the sound velocity now becomes a function of the temperature, thus a function of $\xi$, or to say, a function of $v$ since $v(\xi)$ is a monotonic function of $\xi$ in this sound shell regime. On the other hand, Eq.~\eqref{eq:temperature} still works even beyond the bag EoS, thus we can use it to evaluate temperature profile in the range $\xi<\xi_w$ as
\begin{align}
    T(\xi) = T_- \exp\left[ -\int_{v(\xi)}^{v(\xi_w)}  \gamma^2(v) \mu(\xi(v),v) \mathrm{d}v \right],
\end{align}
where the temperature just behind the bubble wall $T_-$ is fixed by the replacement of $T_+$ as $T_N$ in the definition~\eqref{eq:alphaplusr} of $r$.

\subsubsection{Deflagration}\label{subsec:deflag_wobag}

The deflagration of the weak type $\bar{v}_+<\bar{v}_-\equiv\xi_w$ still occurs for the bubble wall velocity $\xi_w$ moving slower than the sound velocity $c_s^+$ in the symmetric phase in the front of the bubble wall. Since the bag EoS holds in the symmetric phase, the sound velocity $c_s^+$ reads $1/\sqrt{3}$. Therefore, most of the discussions we consider in the last section still could carry over for the deflagration beyond the bag EoS. Requiring the fluid velocity behind the bubble wall to be static,
\begin{align}
    v_- = 0 = \mu(\xi_w,\bar{v}_-),
\end{align}
we recover the relation $r(\xi_w,\alpha_+)$ and consequently $\bar{v}_\pm(\xi_w,\alpha_+)$ from Eq.~\eqref{eq:vbpobag} and Eq.~\eqref{eq:vbmobag}. The shockwave front position follows the same constraints~\eqref{eq:shockfront} as the one with a bag EoS. Then solving the modified EoM, the velocity profile can be obtained. Note that the temperature behind the bubble wall stays as a constant from the zeroth-order result, hence the deviation of the sound velocity from $1/\sqrt{3}$ in the broken phase is also a constant. Some illustrative examples for the iteration results of the modified profiles of the fluid velocity and sound velocity profile are shown in Fig.~\ref{fig:detonav_wobag} and~\ref{fig:detonacs_wobag}. 

Once we have solved the profile of the fluid velocity for the deflagration mode, we can then follow what we have done in the last section to give rise to the profiles of enthalpy and temperature as shown in Fig.~\ref{fig:deflagw_wobag} and Fig.~\ref{fig:deflagT_wobag}.
The enthalpy profile is obtained similarly as in the case with a bag EoS by first applying the junction condition across the shockwave front with
\begin{align}
    w_- = w(\xi_{sh}^-), \quad w_+ = w_N, \quad \bar{v}'_-=\bar{v}'_-(\alpha'_+\equiv0,r'), \quad \bar{v}'_+ = \xi_{sh},
\end{align}
and the enthalpy just behind the shockwave front simply reads
\begin{align}
    w(\xi_{sh}^-) = w_N \frac{\xi_{sh}}{1-\xi_{sh}^2} \frac{1-\mu(\xi_{sh},v(\xi_{sh}))^2}{\mu(\xi_{sh},v(\xi_{sh}))}.
\end{align}
Next we evolve $w( \xi_w )$ backwards from the shockwave front to the bubble wall to render the enthalpy profile in the range $\xi_w < \xi < \xi_{sh}$ as
\begin{align}
    w(\xi) = w(\xi_{sh}^-) \exp\left[ -\int_{v(\xi)}^{v(\xi_{sh})} \left( 1+\frac{1}{c_s^2} \right) \gamma^2(v) \mu(\xi(v),v) \mathrm{d}v \right]
    \label{eq:enthalpy_deflag_wobag}
\end{align}
whhere the sound velocity here can be replaced by $1/\sqrt{3}$ since the integral is done in the symmetric phase. Finally, the enthalpy within the bubble wall is obtained by applying the junction condition across the bubble wall with
\begin{align}
    w_-=w(\xi_w^-), \quad w_+=w(\xi_w^+), \quad \bar{v}_-=\xi_w, \quad 
    \bar{v}_+=\mu(\xi_w,v(\xi_w)),
\end{align}
and the constant enthalpy behind the bubble wall simply reads
\begin{align}
    w(\xi_w^-) = w(\xi_w^+) \frac{1-\xi_w^2}{\xi_w} \frac{\mu(\xi_{w},v(\xi_{w}))}{1-\mu(\xi_{w},v(\xi_{w}))^2},
\end{align}
where $w(\xi_w^+)$ can be evaluated by taking $\xi=\xi_w$ in Eq.~\eqref{eq:enthalpy_deflag_wobag}. The temperature profile between the bubble wall and shockwave front is evolved as
\begin{align}
    T(\xi) = T(\xi_{sh}^-) \exp\left[ -\int_{v(\xi)}^{v(\xi_{sh})}  \gamma^2(v) \mu(\xi(v),v) \mathrm{d}v \right],
\end{align}
where the temperature jump at the shockwave front can be derived from the enthalpy jump as
\begin{align}
    T(\xi_{sh}^-) =  T_N \left( \frac{w(\xi_{sh}^-)}{w_N} \right)^{1/4}
\end{align}
since the bag EoS still applies to the symmetric phase so that 
\begin{align}
    \frac{a_+T_+^4}{a_-T_-^4} = \frac{w_+}{w_-}
\end{align}
still holds across shockwave front, and $w(\xi_{sh}^-)$ has been obtained from Eq.~\eqref{eq:enthalpy_deflag_wobag}. The constant temperature behind the bubble wall is evaluated by the definition of $r$ as
\begin{align}
    T(\xi_w^-) = T(\xi_w^+) \left( \frac{1}{r(\xi_w,\alpha_+)} \frac{a_+}{a_-} \right)^{1/4}.
\end{align}

\subsubsection{Hybrid}

Hybrid expansion occurs when the bubble wall moves faster than the sound velocity but slower than the modified Jouguet velocity, which is a special deflagration mode ($\bar{v}_+<\bar{v}_-$) of Jouguet type ($\bar{v}_-=c_s^-$). For $\xi>\xi_{sh}$, the fluid stays at rest in the bubble center frame, where the shockfront velocity is still constrained by $\xi_{sh}\mu(\xi_{sh},v(\xi_{sh}))=(c_s^+)^2=1/3$. For $\xi_w<\xi<\xi_{sh}$ and $\xi_r<\xi<\xi_{w}$, the velocity profiles are solved from the fluid EoM with the boundary condition $\bar{v}_-=c_s^-$ just behind the bubble wall, where $\xi_r$ is defined by $v(\xi_r) = 0$. Note that $\xi_r=c_s^{(0)}\equiv1/\sqrt{3}$ for a bag EoS but $\xi_r<1/\sqrt{3}$ beyond the bag EoS due to the decrease of the sound velocity in the broken phase. Note also that $v_-=\mu(\xi_w,\bar{v}_-=c_s^-<1/\sqrt{3})>\mu(\xi_w,1/\sqrt{3}=c_s^{(0)}=\bar{v}_-^{(0)})=v_-^{(0)}$, which can be used to check our numerical results. Finally, for $\xi<\xi_r$, the fluid again stays at rest. Some illustrative examples for the iteration results of the modified profiles of the fluid velocity and sound velocity are shown in Fig.~\ref{fig:hybridv_wobag} and~\ref{fig:hybridcs_wobag}. 

Once we have solved the profile of the fluid velocity for the hybrid mode, we can then follow what we have done in the last section to give rise to the profiles of enthalpy and temperature as shown in Fig. ~\ref{fig:hybridw_wobag} and~\ref{fig:hybridT_wobag}, which are evaluated by Eq.~\eqref{eq:enthalpyobag} and Eq.~\eqref{eq:temperature}, respectively. In the front of the shockwave front, both the enthalpy and temperature stay constant given by their asymptotic values far outside of the bubble. Behind the shockwave front but in the front of the bubble wall, the enthalpy and temperature profiles evolve as
\begin{align}
    w(\xi) =& w(\xi_{sh}^-) \exp\left[ -\int_{v(\xi)}^{v(\xi_{sh}^-)} \left( 1+\frac{1}{c_s^2} \right) \gamma^2(v) \mu(\xi(v),v) \mathrm{d}v \right],
    \label{eq:enthalpy_hybrid_wobag} \\
    T(\xi) =& T(\xi_{sh}^-) \exp\left[ -\int_{v(\xi)}^{v(\xi_{sh}^-)}  \gamma^2(v) \mu(\xi(v),v) \mathrm{d}v \right],
\end{align}
where their values just behind the shockwave front are given by the junction conditions as
\begin{align}
    w(\xi_{sh}^-) = w_N \frac{\xi_{sh}}{1-\xi_{sh}^2} \frac{1-\mu(\xi_{sh},v(\xi_{sh}))^2}{\mu(\xi_{sh},v(\xi_{sh}))}, \quad
    T(\xi_{sh}^-) = T_N \left( \frac{w(\xi_{sh}^-)}{w_N} \right)^{1/4}.
\end{align}
Note here that $c_s^2\equiv1/3$ for the symmetric phase in the front of the bubble wall. Behind the bubble wall, the enthalpy and temperature profiles evolve as
\begin{align}
    w(\xi) =& w(\xi_{w}^-) \exp\left[ -\int_{v(\xi)}^{v(\xi_{w}^-)} \left( 1+\frac{1}{c_s^2(T)} \right) \gamma^2(v) \mu(\xi(v),v) \mathrm{d}v \right], \\
    T(\xi) =& T(\xi_{w}^-) \exp\left[ -\int_{v(\xi)}^{v(\xi_{w}^-)}  \gamma^2(v) \mu(\xi(v),v) \mathrm{d}v \right],
\end{align}
where their values just behind the bubble wall are given by the junction condition as
\begin{align}
    w(\xi_w^-) = w(\xi_w^+) \frac{\bar{v}_+}{1-\bar{v}_+^2} \frac{1-(c_s^-)^2}{c_s^-},
    \quad
    T(\xi_w^-) = T(\xi_w^+) \left( \frac{1}{r(\xi_w,\alpha_+)} \frac{a_+}{a_-} \right)^{1/4}.
\end{align}
Note here that $\bar{v}_+$, $w(\xi_w^+)$ and $T(\xi_w^+)$ can be obtained from the velocity, enthalpy and temperature profiles in the region $\xi_w<\xi<\xi_{sh}$, and $v(\xi_{w}^-) = \mu(\xi_w,c_s^-)$ with $c_s^-$ the sound velocity just behind the bubble wall. Finally, evolving the previous evaluation on the fluid velocity to its vanishing point at $\xi=\xi_r$, the enthalpy and temperature profiles in the region $\xi<\xi_r$ again stay constant different from their asymptotic value far in the front.

\subsection{Efficiency factor}\label{subsec:efficiency}

\begin{figure}
    \centering
    \includegraphics[width=0.8\textwidth]{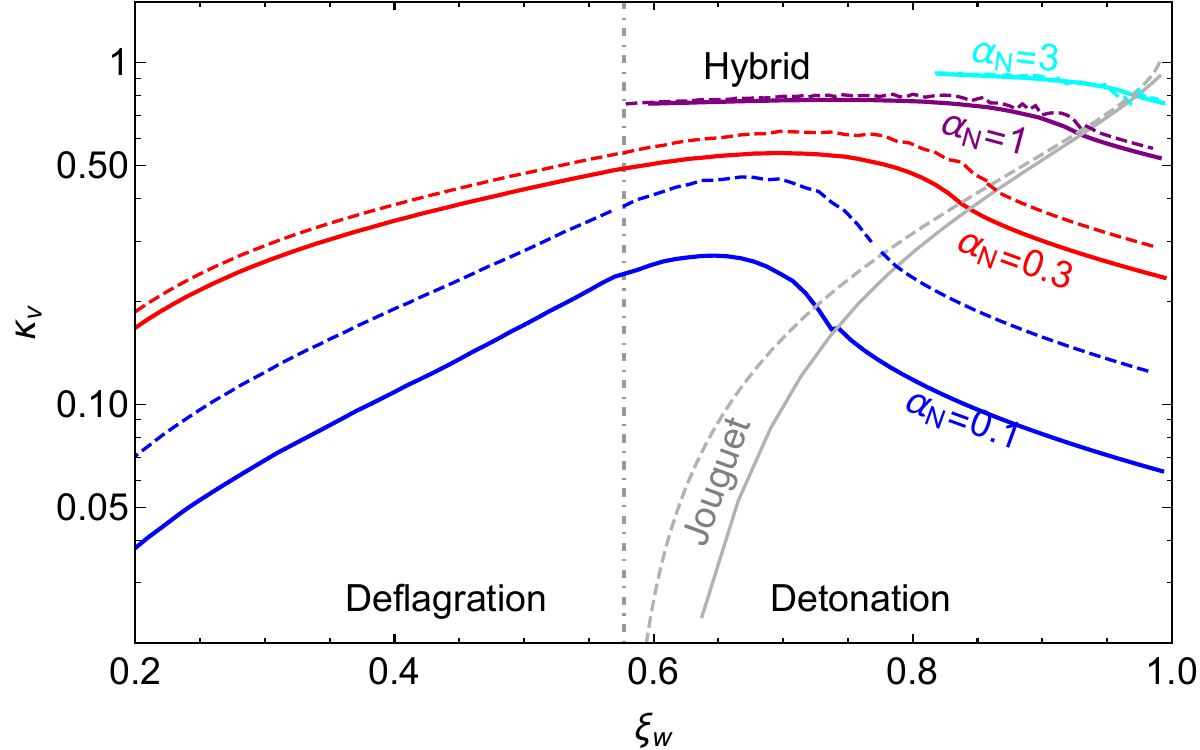}
    \caption{The efficiency factors $\kappa_v$ of bulk fluid motions for a bag EoS (dashed) and our general EoS (solid) are presented with respect to the bubble wall velocity $\xi_w$ given the asymptotic strength factor $\alpha_N$ for a typical set of parameter choices with $a_+/a_-=1.2$, $b_-/(a_- T_N^2)=2/25$, $c_-/(a_- T_N^3)=2/125$, and $T_N = 500$ GeV.}
    \label{fig:Efficiency}
\end{figure}

Having the fluid velocity and enthalpy profiles in hand, we can directly calculate the efficiency factor for the bulk fluid motions, which is defined by the ratio of the bulk fluid kinetic energy with respect to the released vacuum energy as~\cite{Espinosa:2010hh}
\begin{align}
    \kappa_v &= \left. \int w(\xi) v^2 \gamma^2 ~ 4\pi\xi^2 \mathrm{d}\xi \right/
    \left( \frac{4\pi}{3} \Delta\epsilon \cdot \xi_w^3 \right)
    = \frac{3}{\Delta\epsilon \cdot \xi_w^3} \int w(\xi) v^2 \gamma^2 \xi^2 \mathrm{d}\xi
    \nonumber \\
    &= \frac{4}{\alpha_N \xi_w^3} \int_0^1 \frac{w(\xi)}{w_N} v^2 \gamma^2 \xi^2 \mathrm{d}\xi.
    \label{eq:efficiency}
\end{align}
where $\alpha_N$ is the asymptotic value of the strength factor far outside the bubble wall and shockwave front, if any. Since the bag EoS is still valid in the symmetric phase, $\alpha_N$ is thus computed by
\begin{align}\label{eq:alphaNalphap}
\alpha_N=\frac{\Delta\epsilon}{a_NT_N^4}=\frac{a_+T_+^4}{a_NT_N^4}\alpha_+=\frac{w_+}{w_N}\alpha_+
\end{align}
from the enthalpy $w_+/w_N$ and strength factor $\alpha_+$ just in the front of the bubble wall. However, the use of $\alpha_+$ as the input is obscure due to the presence of the shockwave in front of the bubble wall. Therefore, one usually solves  $\alpha_+$ from~\eqref{eq:alphaNalphap} for a given $\alpha_N$, and then expresses the efficiency factor $\kappa_v$ as a function of  $\xi_w$ and $\alpha_N$ for our general EoS with parameters $a,b,c$ normalized appropriately by the asymptotic temperature $T_N$. The numerical results for $\kappa_v$ with a bag EoS (dashed curves) and our general EoS (solide curves) are presented in Fig.~\ref{fig:Efficiency}. Here the modified efficiency factor is evaluated with following strategies: For weak detonation and Jouguet detonation, the velocity and enthalpy profiles are directly solved from the integral-differential EoM. For weak deflagration and hybrid expansion, the velocity and enthalpy profiles come from the first-order iteration results, which are precise enough for our practicle use. 

Although our numerical results are illustrated from a typical set of parameter choice for our general EoS parameters, the same suppression effect of the efficiency factor of bulk fluid motions for our general EoS with respect to the case with a bag EoS is also found in Fig.~2 of Ref.~\cite{Giese:2020znk} for the $\nu$-model with $1/2 = c_s^- < c_s^+ = 1/\sqrt{3}$. We can understand this suppression effect for the three expansion modes separately as follows:

For the detonation mode, solving Eq.~\eqref{eq:vbpobag} with $\bar{v}_+=\xi_w$ results in a larger $\bar{v}_-$ due to the existence of the lower bound on $\alpha_+$ that further leads to smaller $v_-$ than their analogues with a bag EoS. This is consistent with our numerical result in Fig.~\ref{fig:detonav_wobag}, where the velocity just behind the bubble wall indeed becomes slower. Similarly, the enthalpy is also smaller in the broken phase than the case with a bag EoS. Therefore, the fluid behind the bubble wall beyond the bag EoS is less energetic than the case with a bag EoS. Physically, this comes from the modification of the pressure and energy density from Eq.~\eqref{eq:pobag} and~\eqref{eq:rhoobag}, respectively, where, in the case with a bag EoS, we just concentrate on the leading $\mathcal{O}(T^4)$ term but neglect the sub-leading $\mathcal{O}(T^2)$ term that admits a negative contribution. As the summation of the pressure and energy density, the enthalpy also inevitably decreases. Therefore, the bulk fluid gains less energy from the bubble wall expansion than the bag EoS case.

For the deflagration mode, the argument is quite similar. Solving Eq.~\eqref{eq:vbmobag} with $\bar{v}_-=\xi_w$ results in a larger $\bar{v}_+$ due to the existence of the lower bound on $\alpha_+$ that further leads to smaller $v_+$ than their analogues with a bag EoS. Thus, with the same EoM as the bag EoS case (since the sound velocity does not change in the symmetric phase), the modified shockwave front must lie behind the bag one. Therefore, both the velocity and enthalpy become smaller than their analogues with a bag EoS as we can see in Fig.~\ref{fig:deflagv_wobag} and~\ref{fig:deflagw_wobag}. Physically, this comes from the modification of the pressure and energy density in the broken phase. With a decreased pressure in the broken phase, the bubble wall pushes less hard against the bulk fluid beyond bag EoS than the case with a bag EoS. Therefore, the velocity jump at the bubble wall is less energetic and hence the bulk fluid gains less energy from the bubble wall expansion.

For the Hybrid mode, the situation is not as simple as the other two expansion modes. The non-vanishing part of velocity profile contains both rarefaction wave and compressive shockwave. Since the velocity and enthalpy in the shockwave in the front of the bubble wall are several times larger than that in the rarefaction wave, the bulk fluid in the symmetric phase must dominate the contribution to the efficiency factor. Therefore, it is similar to conclude that the efficiency factor is suppressed by just noticing the decrease of the velocity in the shockwave and but neglecting the increase of velocity in the rarefaction wave.

Last but not the least, due to four new parameters ($a_+/a_-, b_-/(a_-T_N^2), c_-/(a_-T_N^3), T_N$) introduced for our general EoS, the numerical fitting formula for $\kappa_v(\xi_w,\alpha_N)$ in a sufficiently large parameter space beyond the bag EoS is not easy to summarized, which will be given in a future study along with the numerical code developed in this paper available for public use. Nevertheless, it is easy to see that the suppression effect for the efficiency factor of bulk fluid motions is less pronounced for stronger FOPT with a larger strength factor. Therefore, the original numerical fitting formula~\cite{Espinosa:2010hh} for the strong FOPT ($\alpha_N\gtrsim1$) in the case with a bag EoS still serves as a good ansatz. 

\section{Conclusions and discussions}\label{sec:con}

The efficiency factor of bulk fluid motions is a key parameter characterizing the gravitational-wave spectrum for the sound wave contribution. The previous estimations for this efficiency factor usually assume a constant sound velocity for a bag EoS model with equal sound velocities or the $\nu$-model with different but still constant sound velocities in the symmetric and broken phases. From the particle physics point of view, both the bag EoS model and $\nu$-model are not realistic enough as a general EoS assumption. Therefore, we propose in this paper to use a more general and realistic EoS by expanding the thermal potential to the higher orders, and then solve the fluid EoM with the iteration method so that the sound velocity profile can be determined consistently with the hydrodynamic solutions. Finally, we directly compare our new estimation for the efficiency factor of bulk fluid motions, which is relatively suppressed with respect to the case with a bag EoS. Nevertheless, for a stronger FOPT with a larger strenght factor, such a suppression effect is less pronounced, hence the previous estimation for the efficiency factor of bulk fluid motions from a bag EoS is still viable as a good approximation. However, this might still depend on the specific particle physics models, where the suppression effect could be significant enough to affect the gravitational-wave spectrum. Future work should be carried out for a numerical fitting formula to cover a sufficiently large parameter space of our general EoS for a practical use.

\acknowledgments
SJW is supported by the National Key Research and Development Program of China Grant  No.2021YFC2203004, No. 2020YFC2201501 and No.2021YFA0718304, 
the National Natural Science Foundation of China Grants 
No. 12105344,
the Key Research Program of the Chinese Academy of Sciences (CAS) Grant No. XDPB15, 
the Key Research Program of Frontier Sciences of CAS, 
and the Science Research Grants from the China Manned Space Project with No. CMS-CSST-2021-B01.

\appendix

\section{Higher order expansion of the effective potential}\label{sec:higher_order}

In this appendix, we present results when expanding the thermal potential beyond the $\mathcal{O}((m_i/T)^3)$ order in the $J_\mathrm{B/F}(m_i^2/T^2)$ function for our general EoS, for example, including the logarithmic terms,
\begin{align}
    \mathcal{F}(\phi,T) = V_{\mathrm{eff}}(\phi,T) \approx V_{0}(\phi) -\frac{1}{3}a T^4 + b T^2 - c T - d \ln\frac{T}{T_0}+C,
\end{align}
where the reference scale $T_0$ can be chosen appropriately to render the constant term 
\begin{align}
C=\frac{1}{32\pi^2}\left(\sum_{i=\mathrm{B}}g_im_i^4\ln\frac{\lambda_BT_0}{m_i}-\sum_{i=\mathrm{F}}g_im_i^4\ln\frac{\lambda_FT_0}{m_i}\right), \quad \lambda_B=4\lambda_F=4\pi e^{3/4-\gamma_E},
\end{align}
with to be zero. The coefficients $b$ and $c$ have been given in Eq.~\eqref{eq:coff}, while the parameter $d$ is given by
\begin{equation}
    \begin{aligned}
        d = \frac{1}{32\pi^2} \left( -\sum_{i=\mathrm{B}}g_i m_i^4 + \sum_{j=\mathrm{F}}g_j m_j^4 \right) . \label{eq:coffd}
    \end{aligned}
\end{equation}
where the signs in front of the boson/fermion contributions to $d$ are chosen in such a way so that $d$ is positive for  SM due to the heavy mass of top quarks. However, for many particle physics models beyond SM (especially in SUSY), there could be a lot of new particles contributing to $d$ so that the sign of $d$ would not necessarily be positive. Therefore, it is not easy to determine the sign of $d$ for a general model beyond SM. In this appendix, we will simply assume a positive $d$ for illustration.

\begin{figure}[ht]
    \centering
    \subfigure[$v(\xi)$ for detonation]{
    \includegraphics[width=0.47\textwidth]{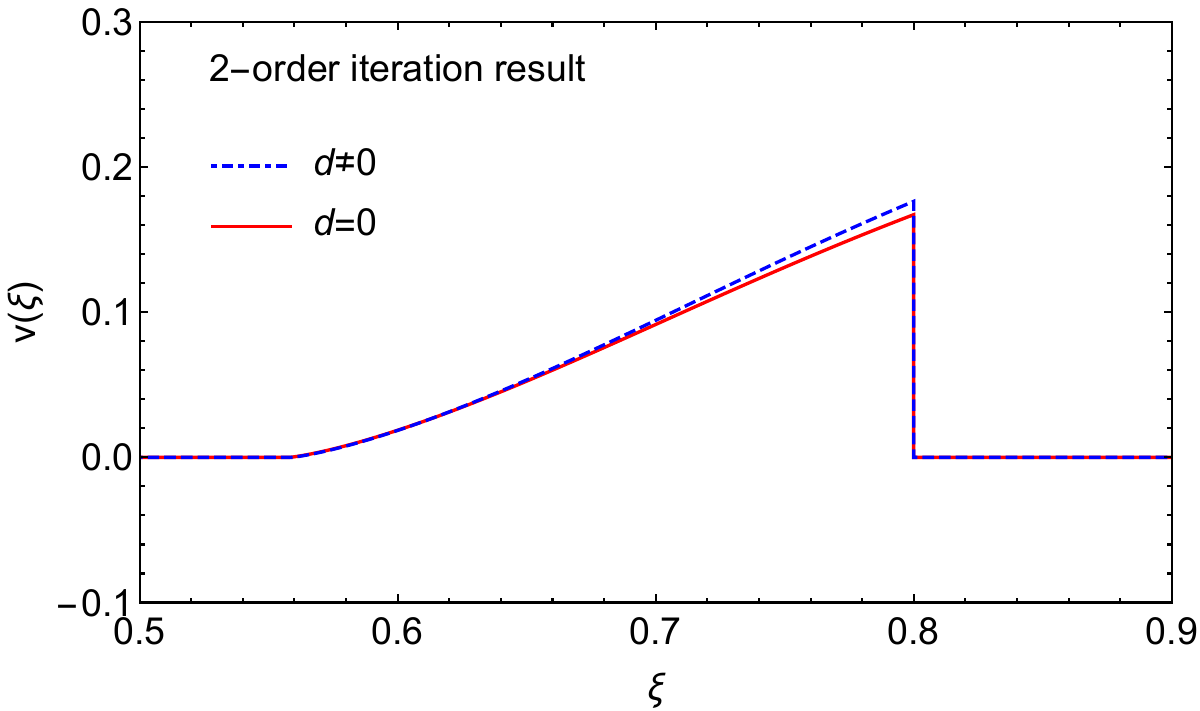}
    }
    \subfigure[$w(\xi)/w_N$ for detonation]{
    \includegraphics[width=0.47\textwidth]{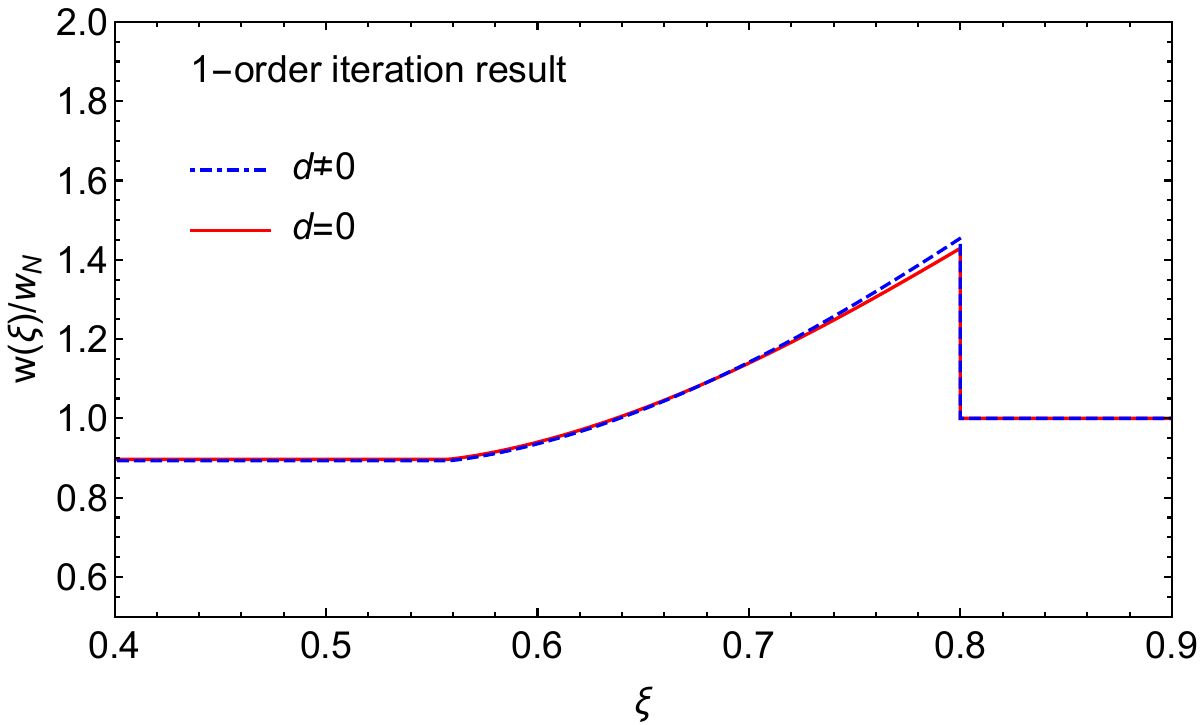}
    }
    \subfigure[$v(\xi)$ for hybrid]{
    \includegraphics[width=0.47\textwidth]{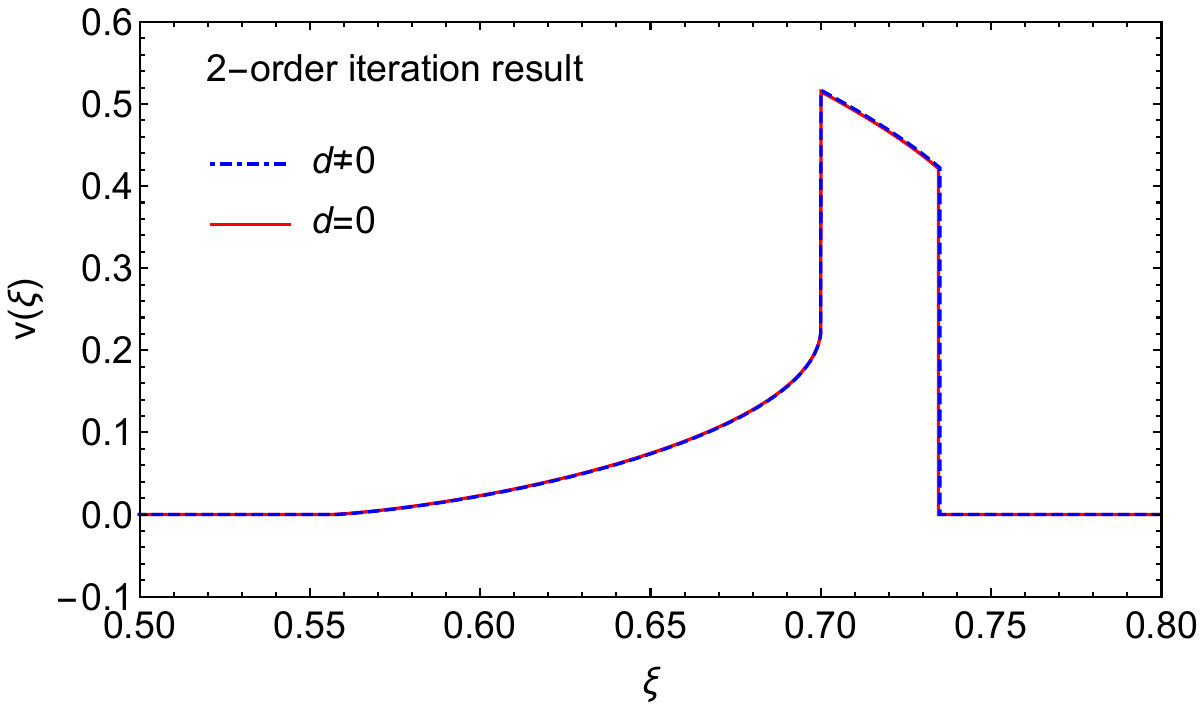}
    }
    \subfigure[$w(\xi)/w_N$ for hybrid]{
    \includegraphics[width=0.47\textwidth]{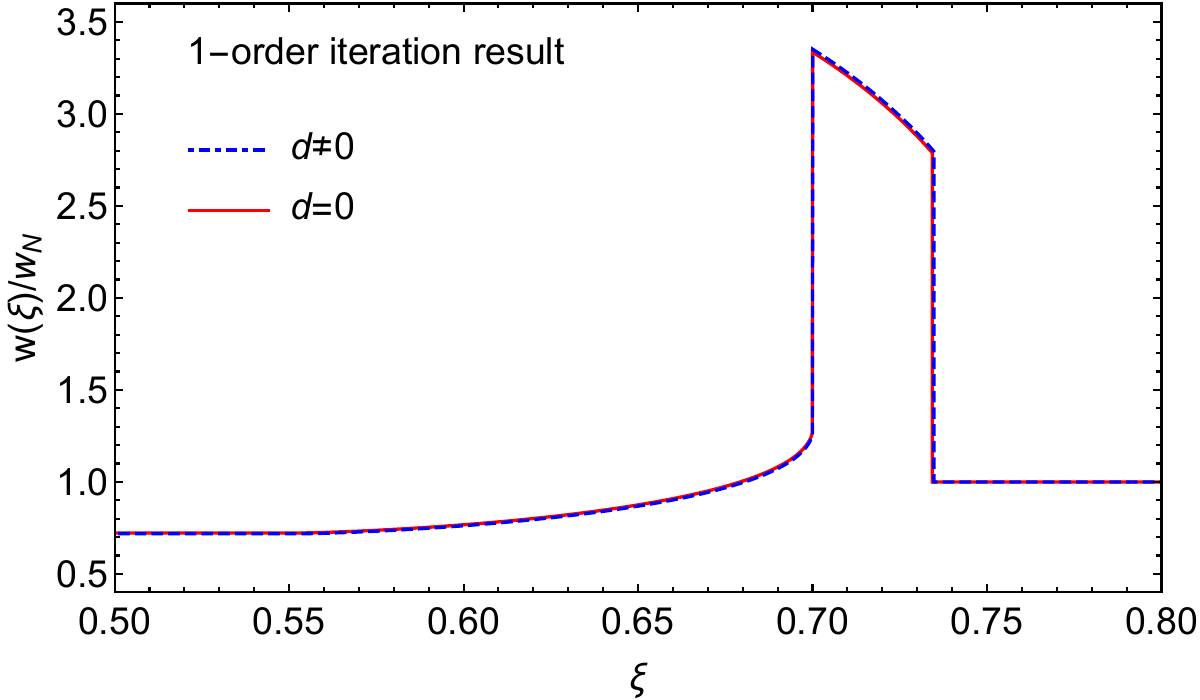}
    }
    \subfigure[$v(\xi)$ for deflagration]{
    \includegraphics[width=0.47\textwidth]{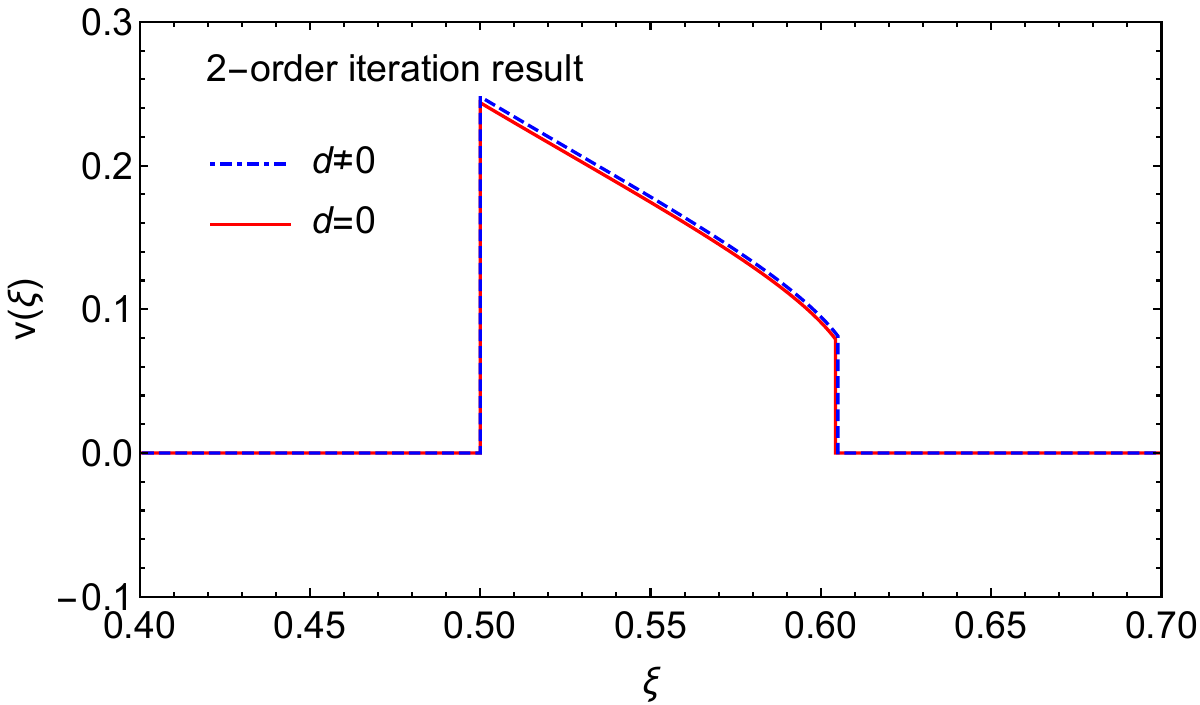}
    }
    \subfigure[$w(\xi)/w_N$ for deflagration]{
    \includegraphics[width=0.47\textwidth]{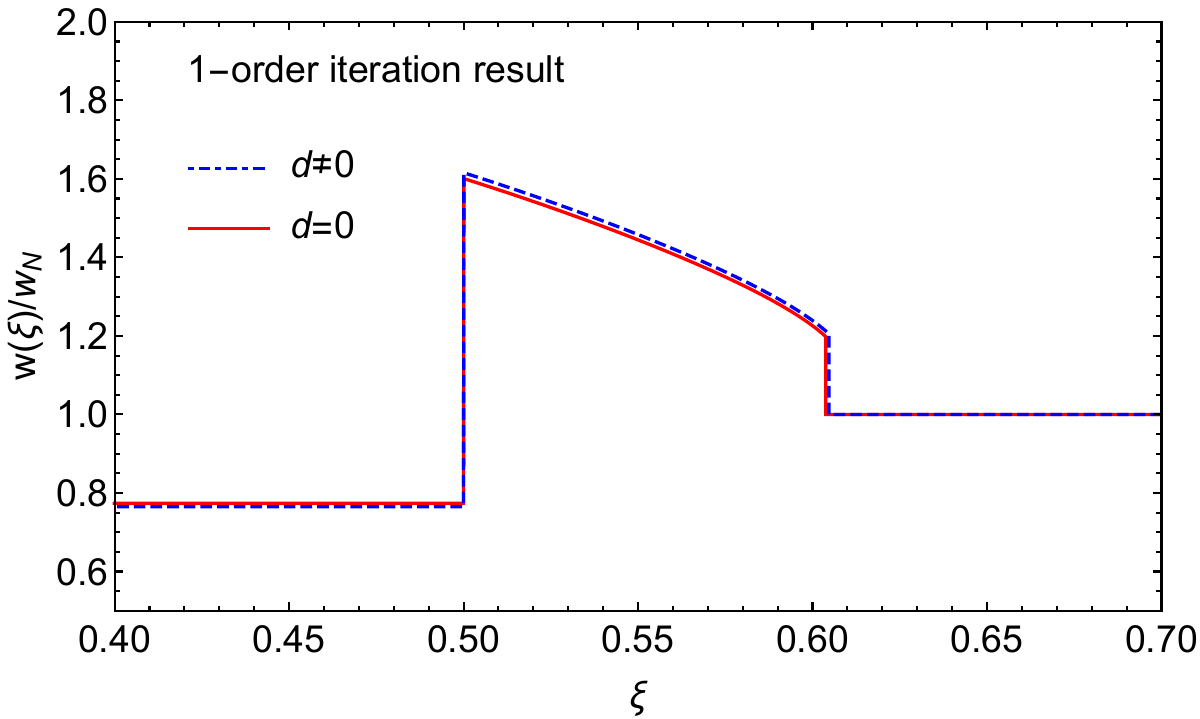}
    }
    \caption{The comparison of numerical results for the profiles of the fluid velocity (left column) and enthalpy (right column) with $d_-=0$ (red solid) and $d_-/(a_- T_N^4)=2/625$ (blue dashed). Here we show the second-order iteration results for the fluid velocity profiles and first-order iteration results for the enthalpy profiles. Other parameters $\alpha_+=0.1$, $a_+/a_-=1.2$, $b_-/(a_- T_N^2)=2/25$, $c_-/(a_- T_N^3)=2/125$, and $T_N = 500$ GeV  are fixed to be the same for all cases, and $\xi_w = 0.8$, $0.7$, $0.5$ for detonation (top row), hybrid (middle row) and deflagration (bottom row), respectively. The bag EoS is still assumed in the symmetric phase for all cases.}
    \label{fig:comparison_obagA}
\end{figure}

Using the $V_{\mathrm{eff}}$ above, we can directly get the pressure and energy density of forms
\begin{align}
    p = -V_{0}(\phi) + \frac{1}{3}a T^4 - b T^2 + cT + d\ln\frac{T}{T_0}, \label{eq:pobagA}\\
    \rho = V_{0}(\phi) + a T^4 -b T^2 - d\ln\frac{T}{e T_0} . \label{eq:rhoobagA}
\end{align}
The sound velocity will also acquire a deviation from $1/3$, and in the high temperature limit to $T^{-4}$ order, it reads
\begin{align}
    c_s^2 = \frac{1}{3} - \frac{b}{3a T^2} + \frac{c}{4a T^3} - \frac{b^2 - 2ad}{6a^2 T^4} + \mathcal{O}(T^{-5}) . \label{eq:cs2obagA}
\end{align}
Again we assume that the bag EoS is still kept in the symmetric phase outside the bubble wall, thus $b_+=c_+=d_+0$. Apply this new general EoS Eq.~\eqref{eq:pobagA} and Eq.~\eqref{eq:rhoobagA} to the junction conditions, we can reexpress $\bar{v}_\pm$ as
\begin{align}
\bar{v}_+&=\sqrt{\frac{F(T_-)-(1-3\alpha_+)r}{G(T_-)-3(1+\alpha_+)r}\cdot
\frac{G(T_-)+(1-3\alpha_+)r}{F(T_-)+3(1+\alpha_+)r}}, \label{eq:vbpobagA} \\
\bar{v}_-&=\sqrt{\left.\frac{F(T_-)-(1-3\alpha_+)r}{G(T_-)-3(1+\alpha_+)r}
\right/\frac{G(T_-)+(1-3\alpha_+)r}{F(T_-)+3(1+\alpha_+)r}}, \label{eq:vbmobagA}
\end{align}
with $T_\pm$ to be the temperature right in front and back of the bubble wall, and the abbreviations
\begin{align}
\alpha_+=\frac{\Delta\epsilon}{a_+T_+^4}=\frac{4\Delta\epsilon}{3w_+}, \quad r=\frac{a_+ T_+^4}{a_-T_-^4}, \\
F(T) = 1- \frac{3b_-}{a_-} \frac{1}{T^2} + \frac{3c_-}{a_-} \frac{1}{T^3} + \frac{3d_-}{a_-} \frac{1}{T^4}\ln\frac{T}{T_0}, \label{eq:FTA} \\
G(T) = 3- \frac{3b_-}{a_-} \frac{1}{T^2} + \frac{3d_-}{a_-} \frac{1}{T^4} \ln\frac{T}{eT_0}.  \label{eq:GTA}
\end{align}
Now with the modified fluid EoMs and junction conditions, one can numerically solve for the fluid velocity, enthalpy and temperature profiles following the same method detailed in Section~\ref{sec:wobag}. The comparison of the numerical results of the fluid velocity and enthalpy profiles with and without $d$ parameter are showed in Fig.~\ref{fig:comparison_obagA}, which is negligibly small for our typical choice of parameters.

\section{Corrections from particles with $m_i\gtrsim T$}\label{sec:Massive_particles}

Our EoS ansatz~\eqref{eq:VeffbeyondbagEoS} only applies to the case when the particle masses in the broken phase are either smaller than the phase transition temperature $m_i<T_N$ or much heavier than the phase transition temperature $m_i\gg T_N$. When some particles receive masses slightly heavier than the phase transition temperature $m_i\gtrsim T_N$, the low-temperature expansion~\eqref{eq:JBF_lowT} is not valid anymore and its the exponential suppression factor could be comparable to those light particles in the high-temperature expansion~\eqref{eq:VeffbeyondbagEoS}. In this special case, we should work out the conditions when these slightly heavier particles cannot be simply neglected in the effective potential
\begin{align}
    V_{\mathrm{eff}}(\phi,T) \approx V_{0}(\phi) -\frac{1}{3}a T^4 + b T^2  - c T + f(T)T^{5/2},
\end{align}
where the parameters $a$, $b$, and $c$ in the truncated high-temperature expansion are evaluated for all light particles with $m_i<T_N$, while the $f(T)$ term is evaluated over all heavy particles with $m_i\geq T_N$ but dominated by those slightly heavier particles with $m_i\gtrsim T_N$,
\begin{align}
    f(T) &= \left(\frac{1}{32\pi^7}\right)^{1/2} \sum_{m_i\ge T_N}g_i m_i^{3/2}e^{-m_i/T}\left(1+\mathcal{O}(\frac{T}{m_i})\right).
\end{align}
Let us compare this $f(T)T^{5/2}$ term to the terms involving with $a$, $b$, and $c$, respectively.

First, the $(1/3)aT_N^4$ term contains all light particles with $m_i<T_N$, while the $f(T_N)T_N^{5/2}$ term is dominated by the slightly heavier particles with $m_i\gtrsim T_N$, thus their ratio reads
\begin{align}
\frac{\frac13aT_N^4}{f(T_N)T_N^{5/2}}
&=\frac{\frac{\pi^2}{90}\sum\limits_{m_i<T_N}\left(g_{i,B}+\frac78g_{i,F}\right)T_N^4}{\left(\frac{1}{32\pi^7}\right)^{1/2}\sum\limits_{m_i\geq T_N}g_i\left(\frac{m_i}{T_N}\right)^\frac32e^{-\frac{m_i}{T_N}}T_N^4}
\simeq\frac{10^{-1}\sum\limits_{m_i<T_N}\left(g_{i,B}+\frac78g_{i,F}\right)}{10^{-3}\sum\limits_{m_i\gtrsim T_N}g_i\left(\frac{m_i}{T_N}\right)^\frac32e^{-\frac{m_i}{T_N}}} \nonumber \\
&\lesssim\frac{10^{-1}g_\mathrm{eff}}{10^{-4}\sum\limits_{m_i\gtrsim T_N}g_i}<1\Rightarrow \sum\limits_{m_i\gtrsim T_N}g_i>10^3g_\mathrm{eff},
\end{align}
where $e^{-m_i/T_N}\sim\mathcal{O}(10^{-1})$ is estimated for $m_i\gtrsim T_N$. Therefore, the $f(T_N)T_N^{5/2}$ term would dominate over the leading $(1/3)aT_N^4$ term if the number of the degrees of freedom from the slightly heavier particles is thousand times larger than the total effective degrees of freedom of all light particles, which is highly unlikely in current model buildings on the market.

Second, the $bT_N^2$ term is dominated by those slightly light particles with $m_i\lesssim T_N$, hence the ratio between the $bT_N^2$ term and the $f(T_N)T_N^{5/2}$ term could be estimated as
\begin{align}
\frac{bT_N^2}{f(T_N)T_N^{5/2}}
&=\frac{\frac{1}{24}\left(\sum\limits_{m_{i,B}<T_N}g_{i,B}\frac{m_{i,B}^2}{T_N^2}+\frac12\sum\limits_{m_{i,F}<T_N}g_{i,F}\frac{m_{i,F}^2}{T_N^2}\right)T_N^4}{\left(\frac{1}{32\pi^7}\right)^{1/2}\sum\limits_{m_i\geq T_N}g_i\left(\frac{m_i}{T_N}\right)^\frac32e^{-\frac{m_i}{T_N}}T_N^4}
\simeq\frac{10^{-2}\sum\limits_{m_i\lesssim T_N}g_i\left(\frac{m_i}{T_N}\right)^2}{10^{-3}\sum\limits_{m_i\gtrsim T_N}g_i\left(\frac{m_i}{T_N}\right)^\frac32e^{-\frac{m_i}{T_N}}}\nonumber\\
&\lesssim\frac{10^{-2}\sum\limits_{m_i\lesssim T_N}g_i}{10^{-4}\sum\limits_{m_i\gtrsim T_N}g_i}<1\Rightarrow \sum\limits_{m_i\gtrsim T_N}g_i>10^2\sum\limits_{m_i\lesssim T_N}g_i.
\end{align}
Therefore, the $f(T_N)T_N^{5/2}$ term would dominate over the next-leading $bT_N^2$ term if the number of the degrees of freedom from the slightly heavier particles is hundred times larger than the number of the degrees of freedom of from the slightly light particles, which is also highly unlikely in current model buildings on the market.

Third, the $cT_N$ term is also dominated by those slightly light bosons with $m_i\lesssim T_N$, hence the ratio between the $cT_N$ term and the $f(T_N)T_N^{5/2}$ term could be estimated as
\begin{align}
\frac{cT_N}{f(T_N)T_N^{5/2}}
&=\frac{\frac{1}{12\pi}\sum\limits_{m_i<T_N}g_i\left(\frac{m_i}{T_N}\right)^3T_N^4}{\left(\frac{1}{32\pi^7}\right)^{1/2}\sum\limits_{m_i\geq T_N}g_i\left(\frac{m_i}{T_N}\right)^\frac32e^{-\frac{m_i}{T_N}}T_N^4}
\simeq\frac{10^{-2}\sum\limits_{m_i\lesssim T_N}g_i\left(\frac{m_i}{T_N}\right)^3}{10^{-3}\sum\limits_{m_i\gtrsim T_N}g_i\left(\frac{m_i}{T_N}\right)^\frac32e^{-\frac{m_i}{T_N}}}\nonumber\\
&\lesssim\frac{10^{-2}\sum\limits_{m_i\lesssim T_N}g_i}{10^{-4}\sum\limits_{m_i\gtrsim T_N}g_i}<1\Rightarrow \sum\limits_{m_i\gtrsim T_N}g_i>10^2\sum\limits_{m_i\lesssim T_N}g_i.
\end{align}
Therefore, the $f(T_N)T_N^{5/2}$ term would dominate over the next-leading $cT_N$ term if the number of the degrees of freedom from the slightly heavier particles is hundred times larger than the number of the degrees of freedom from the slightly light bosons, which is still highly unlikely in current model buildings on the market.

In a short summary, as long as the number of the degrees of freedom from those slightly heavier particles is not hundred times larger than the number of the degrees of freedom from the slightly light particles, the contribution from the low-temperature expansion of those slightly heavier particles to the effective potential could still be negligible compared to those light particles in our truncated high-temperature expansion. This is usually the case in the current model buildings on the market, for example, the SMEFT with a dimension-six operators $|H|^6/\Lambda^2$~\cite{Ellis:2018mja,Postma:2020toi}.  For a phase transition temperature $T_N=100$ GeV corresponding to a cut-off scale $\Lambda\approx 660$ GeV, the Higgs boson and top quark are slightly heavier than $T_N$, whose contribution in the effective potential can be directly estimated to be smaller than the other terms in our truncated expansion as seen from
\begin{align}
\frac13aT_N^4&=\frac{\pi^2}{90}\times106.75\times T_N^4=1.17\times10^9\,\mathrm{GeV}^4,\\
bT_N^2&=\frac{1}{24}\sum_{i=W,Z}g_im_i^2 \times T_N^2=1.16\times10^7\,\mathrm{GeV}^4,\\
cT_N&=\frac{1}{12\pi}\sum_{i=W,Z}g_im_i^3 \times T_N=6.24\times10^6\,\mathrm{GeV}^4,\\
f(T_N)T_N^{5/2}&=\frac{T_N^4}{2\pi^2}\sum_{i=h,t}g_i\left(\frac{m_i}{2\pi T_N}\right)^{\frac32}e^{-m_i/T_N}=1.68\times10^6\,\mathrm{GeV}^4.
\end{align}

\section{Comparison to the $\nu$-model}\label{sec:compare_with_nu}

\begin{figure}[t]
    \centering
    \subfigure[Detonation]{
    \includegraphics[width=0.3\textwidth]{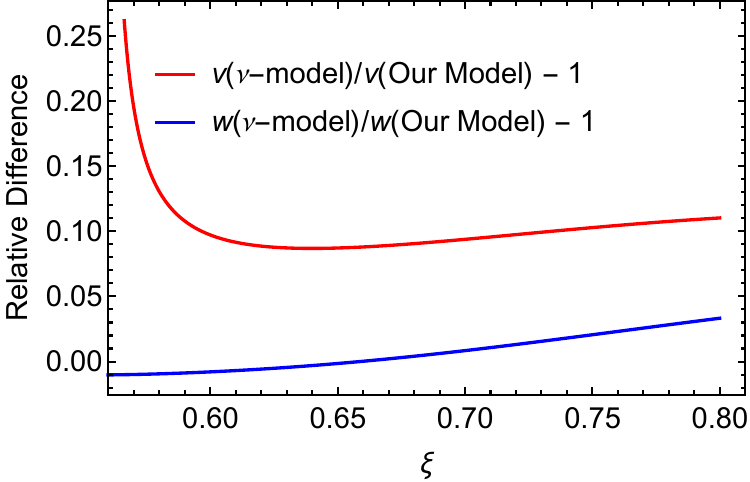}\label{fig:comparison_nu_detona}
    }
    \subfigure[Hybrid]{
    \includegraphics[width=0.3\textwidth]{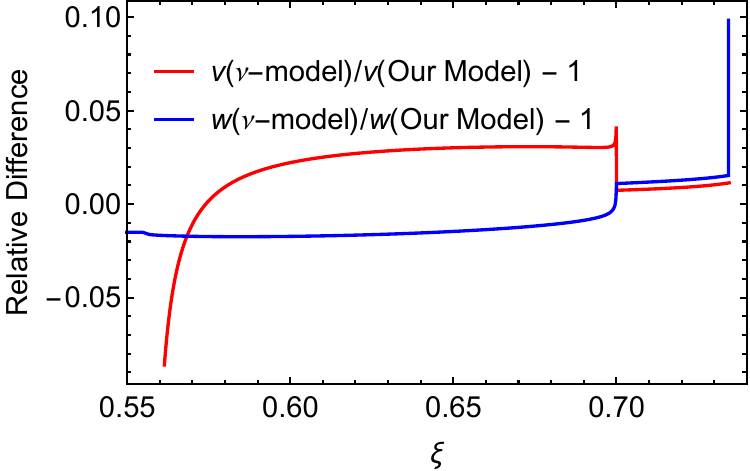}\label{fig:comparison_nu_hybrid}
    }
    \subfigure[Deflagration]{
    \includegraphics[width=0.3\textwidth]{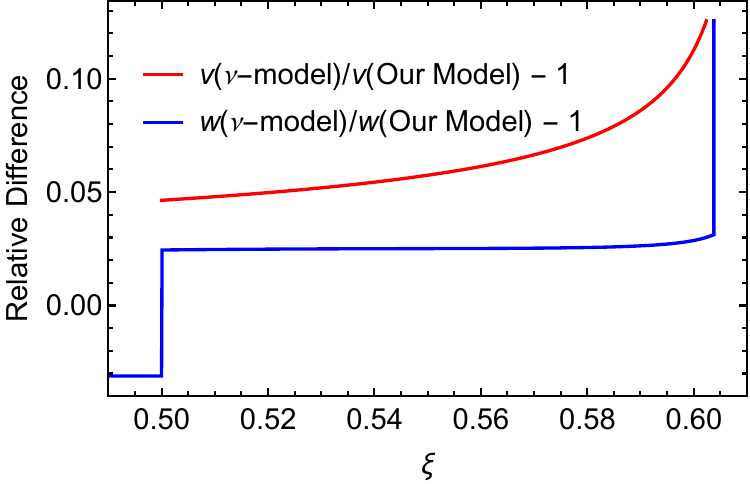}\label{fig:comparison_nu_deflag}
    }
    \caption{The relative difference in the profiles of the fluid velocity and enthalpy within the sound-shell part between the $\nu$-model and our model with the sound velocity deep inside the bubble fixed at $1/(\nu-1)=0.31$ for our typical choices of the parameter values $\alpha_+=0.1$, $a_+/a_-=1.2$, $b_-/(a_- T_N^2)=2/25$, $c_-/(a_- T_N^3)=2/125$, and $T_N = 500$ GeV for the bubble expansions of detonation, hybrid and deflagration types with $\xi_w = 0.8$, $0.7$, $0.5$, respectively.}
    \label{fig:comparison_nu}
\end{figure}

In section \ref{subsec:expansion_wobag}, we have compared the results from our EoS modeling to the case with a bag EoS, which is inputted as the zeroth-order profile when solving the fluid EoM with the iteration method. As one can see in Fig.~\ref{fig:velocity_profiles_obag}, for the bubble expansion of deflagration type, the sound velocity profile converges to the case with a EoS from the $\nu$-model. For the bubble expansion of detonation and hybrid types, the sound velocity profile behind the bubble wall gently approaches to a constant sound velocity, which could be similar to the $\nu$-model. Therefore, it is intriguing to compare our EoS model to the $\nu$-model. To make a fair comparison, we should fix the sound velocity deep inside the bubble to be the same constant value. For our typical choice of the parameter values in the truncated expansion of the effective potential, this constant can be fixed approximately at $1/(\nu-1)=0.31$ for the $\nu$-model. The relative differences in the profiles of the fluid velocity and enthalpy within the sound shell (the non-vanishing part of the fluid velocity) are shown in Fig.~\ref{fig:comparison_nu}, which admits no more than $10\%$ in the relative difference for most of the regimes.

\bibliographystyle{JHEP}
\bibliography{ref}

\end{document}